\documentclass[10pt,english,a4paper]{article}

\usepackage{ifpdf}
\usepackage{multido}
\usepackage{ifthen}
\usepackage{color}

\ifpdf
\usepackage[pdftex]{graphicx}
\else
\usepackage[dvips]{graphicx}
\fi

\usepackage{float,floatflt,afterpage}
\usepackage{wrapfig}
\usepackage{placeins}

\usepackage{etex}
\usepackage{tikz}
\usetikzlibrary{snakes}

\usepackage[arrow,matrix,curve]{xy} \usepackage{subfigure}
\graphicspath{{./figures/}}

\usepackage[latin1]{inputenc}
\usepackage[T1]{fontenc}

\usepackage[centertags,intlimits]{amsmath}
\usepackage{amsthm}
\usepackage{amssymb}
\usepackage{mathtools}
\usepackage{mathrsfs}

\theoremstyle{break}
\allowdisplaybreaks[1]

\usepackage{exscale}
\usepackage{dsfont}

\usepackage[numbers,square,sort&compress]{natbib}

\ifpdf
\usepackage[pdftex,
  bookmarks,
  bookmarksopen=true,
  bookmarksnumbered=true,
  backref,
  linkcolor=blue,
  urlcolor=blue,
  pdfauthor={Andreas Fischle},
  pdftitle={Griolis theorem with weights and the relaxed-polar
    mechanism of optimal Cosserat rotations}
  hyperfigures=true,
  pdfpagelabels,
  pagebackref,
  hypertexnames=true,
  plainpages=false,
  naturalnames]{hyperref}\else
\usepackage[dvips,breaklinks]{hyperref}                                         \usepackage{breakurl}
\fi

\usepackage{hypernat}
\usepackage{url}
\usepackage{breakurl}

\makeatletter
\def\url@leostyle{  \@ifundefined{selectfont}{\def\UrlFont{\sf}}{\def\UrlFont{\small\ttfamily}}}
\makeatother
\usepackage{enumerate}
\usepackage{verbatim}        \usepackage{lscape}          \usepackage{array,longtable} \usepackage{listings}        \usepackage{caption}

\usepackage{makeidx}

                                                                                                                                                                                                                                                                                                            \def\and{{\rm and}}            \def\det#1{{\rm det}\,{#1}}            
                                                  
\def\skew{{\rm skew}}                                  
\newfont{\Sf}{cmssbx10 scaled 2074}
\newbox{\assem}
\savebox{\assem}(15,2)[bl]{
\begin{normalsize}
\unitlength=1.0mm
\begin{picture}(6.00,0.00)
\put(.0,1.0){\makebox(0,0)[cc]{$\cup$}}
\put(.0,-1.0){\makebox(0,0)[cc]{$\mbox{\scriptsize \it n=1}$}}
\put(.0,4.0){\makebox(0,0)[cc] {\scriptsize $\it n_{\mbox{\tiny\it elm}}$}}
\end{picture}
\end{normalsize}}

\newbox{\asse}
\savebox{\asse}(15,2)[bl]{
\begin{normalsize}
\unitlength=1.0mm
\begin{picture}(8.00,0.00)
\put(.0,2.0){\makebox(0,0)[cc]{\Sf A}}
\end{picture}
\end{normalsize}}

                                                        \def\sqtwo3{{\textstyle {\sqrt{2 \over 3}}}}   \newcommand{\IP}{{\rm I\kern-.18em P}}           \newcommand{\II}{{\rm I\kern-.18em I}}           \newcommand{\IF}{{\rm I\kern-.25em F}}           \newcommand{\IE}{{\rm I\kern-.25em E}}           \def\IR{{\rm I\kern-.15em R}}

\newcommand{\ia}{{\rm\kern.24em                     \vrule width.02em height0.9ex depth-.05ex
   \kern-.26em a}}
\newcommand{\ic}{{\rm\kern.24em                     \vrule width.02em height0.9ex depth-.05ex
   \kern-.26em c}}
\newcommand{\IA}{{\rm\kern.22em                      \vrule width.02em
        height0.5ex depth 0ex
    \kern-.24em A}}
\newcommand{\IC}{{\rm\kern.24em                     \vrule width.02em height1.4ex depth-.05ex
   \kern-.26em C}}

\DeclareMathOperator{\sign}{sign}
\DeclareMathOperator{\dist}{dist}

\newcommand{\malpha}{\alpha}
\newcommand{\palpha}{\alpha_\text{p}}
\renewcommand{\epsilon}{\varepsilon}

\newcommand{\norm}[1]{\|#1\|}
\newcommand{\abs}[1]{\left| #1 \right|}

\newtheorem{lem}{Lemma}[section]

\newtheorem{rem}[lem]{Remark}
\newtheorem{defi}[lem]{Definition}

\newtheorem{theo}[lem]{Theorem}

\newtheorem{cor}[lem]{Corollary}

\newtheorem{prob}[lem]{Problem}
\newtheorem{propo}[lem]{Proposition}

\newcommand{\leref}[1]{Lemma \ref{#1}}

\newcommand{\theref}[1]{Theorem \ref{#1}}

\newcommand{\coref}[1]{Corollary \ref{#1}}

\newcommand{\remref}[1]{Remark \ref{#1}}
\newcommand{\deref}[1]{Definition \ref{#1}}

\newcommand{\probref}[1]{Problem \ref{#1}}
\newcommand{\proporef}[1]{Proposition \ref{#1}}

\newcommand{\R}{\mathbb{R}}

\newcommand{\C}{\mathbb{C}}

\renewcommand{\S}{\mathbb{S}}
\DeclareMathOperator{\SL}{SL}
\DeclareMathOperator{\GL}{GL}
\DeclareMathOperator{\SO}{SO}
\let\O\relax
\DeclareMathOperator{\O}{O}

\DeclareMathOperator{\skewop}{skew}
\renewcommand{\skew}{\skewop}
\DeclareMathOperator{\diag}{diag}
\DeclareMathOperator{\sym}{sym}
\DeclareMathOperator{\Tr}{tr}

\DeclareMathOperator{\so}{\mathfrak{so}}

\DeclareMathOperator{\polar}{R_{\rm p}}

\newcommand{\Sym}{ {\rm{Sym}} }
\newcommand{\Psym}{ {\rm{PSym}} }

\newcommand{\id}{{\boldsymbol{\mathds{1}}}}
 
\DeclareMathOperator{\Det}{det}

\renewcommand{\det}[1]{ {\Det[{#1}]} }
\newcommand{\tr}[1]{ {\Tr \left[{#1}\right]} }

\newcommand{\secref}[1]{Section \ref{#1}}
\newcommand{\figref}[1]{Figure \ref{#1}}

\definecolor{orange}{rgb}{1.0,0.5,0}

\DeclareMathOperator{\Reals}{\mathbb{R}}
\renewcommand{\R}{\Reals}

\DeclareMathOperator{\argminmathop}{arg\,min}
\newcommand{\argmin}[2]{\mathchoice{\underset{#1}{\argminmathop}\, {#2}}{\argminmathop_{#1}\, {#2}}{}{}}

\newcommand{\setdef}[2]{\lbrace #1 \;\vert\; #2\rbrace}

\DeclareMathOperator{\spanop}{span}

\newcommand{\vspan}[1]{\spanop\left(\left\{ #1 \right\}\right)}

\newcommand{\ddtat}[2]{\frac{\rm d}{\rm dt}\,#1\big{\vert}_{t = #2}}

\newcommand{\hsnorm}[1]{\left\lVert #1 \right\rVert}

\DeclareMathOperator{\Diag}{Diag}

\DeclareMathOperator{\RPosZ}{\sideset{}{_0^+}\Reals}

\newcommand{\eqdef}{\coloneqq}
\newcommand{\defeq}{\eqqcolon}

\newcommand{\isequivto}{\Longleftrightarrow}

\newcommand{\mstretch}{\overline{U}}

\DeclareMathOperator{\rpolar}{rpolar}

\newcommand{\ump}{u_{\rm mmp}}
\newcommand{\smp}{s_{\rm mmp}}

\DeclareMathOperator{\sradmm}{\rho_{\mu,\,\mu_c}}
\DeclareMathOperator{\sradmmdef}{\frac{2\,\mu}{\mu\,-\,\mu_c}}

\DeclareMathOperator{\wmm}{W_{\mu,\mu_c}}
\DeclareMathOperator{\wsym}{W_{1,0}}

\DeclareMathOperator{\wsymred}{W^{\rm red}_{1,0}}

\newcommand{\domc}{\mathrm{D}^\mathrm{C}}
\newcommand{\domn}{\mathrm{D}^\mathrm{NC}}

\newcommand{\removedAccepted}[1]{}

\newcommand{\countres}{
  \setcounter{equation}{0}
  \setcounter{figure}{0}
  \setcounter{table}{0}
}

                     \renewcommand{\baselinestretch}{1.0}          

\makeatletter

\renewcommand{\itemize}{  \ifnum \@itemdepth >\thr@@\@toodeep\else
    \advance\@itemdepth\@ne
    \edef\@itemitem{labelitem\romannumeral\the\@itemdepth}    \expandafter
    \list
      \csname\@itemitem\endcsname
      {\def\makelabel##1{\hss\llap{##1}}        \topsep=.8ex\itemsep=-.2ex}  \fi}

\renewcommand\section{\@startsection {section}{1}{\z@}  {-3.5ex \@plus -1ex \@minus -.2ex}  {2.3ex \@plus.2ex}  {\boldmath\normalfont\Large\bfseries}}
\renewcommand\subsection{\@startsection{subsection}{2}{\z@}  {-3.25ex\@plus -1ex \@minus -.2ex}  {1.5ex \@plus .2ex}  {\boldmath\normalfont\large\bfseries}}
\renewcommand\subsubsection{\@startsection{subsubsection}{3}{\z@}  {-3.25ex\@plus -1ex \@minus -.2ex}  {1.5ex \@plus .2ex}  {\boldmath\normalfont\normalsize\bfseries}}
\renewcommand\paragraph{\@startsection{paragraph}{4}{\z@}  {3.25ex \@plus1ex \@minus.2ex}  {-1em}  {\boldmath\normalfont\normalsize\bfseries}}
\renewcommand\subparagraph{\@startsection{subparagraph}{5}{\parindent}  {3.25ex \@plus1ex \@minus .2ex}  {-1em}  {\boldmath\normalfont\normalsize\bfseries}}

\makeatother
 
\makeatletter
\renewcommand*{\@fnsymbol}[1]{\ensuremath{\ifcase#1\or 1\or 2\or 3\or
    \mathsection\or \mathparagraph\or \|\or **\or \dagger\dagger
    \or \ddagger\ddagger \else\@ctrerr\fi}}
\makeatother

\title{Grioli's Theorem with weights and the relaxed-polar
       mechanism of optimal Cosserat rotations}

\author{Andreas Fischle
\!\thanks{Corresponding author: Andreas Fischle,
Institut f\"ur Numerische Mathematik,
TU Dresden,
Zellescher Weg 12-14,
01069 Dresden,
Germany,
email: andreas.fischle@tu-dresden.de}
\quad and \quad
Patrizio Neff
\!\thanks{Patrizio Neff,
Head of Lehrstuhl f\"{u}r Nichtlineare Analysis und Modellierung,
Fakult\"{a}t f\"{u}r Mathematik,
Universit\"{a}t Duisburg-Essen,
Thea-Leymann Str. 9,
45127 Essen,
Germany,
email: patrizio.neff@uni-due.de}}

\begin{document}
\selectfont
\maketitle

\pagenumbering{arabic}

\begin{center}\textbf{Abstract}\end{center}
\begin{center}
  \begin{minipage}{0.95\textwidth}
    Let $F \in \GL^+(3)$ and consider the right polar decomposition
    \mbox{$F = \polar(F)\cdot U$} into an orthogonal factor
    $\polar(F) \in \SO(3)$ and a symmetric, positive definite
    factor $U = \sqrt{F^TF} \in \Psym(3)$. In 1940 Giuseppe Grioli
    proved that
    \begin{align*}
      \argmin{R\,\in\,\SO(3)}{\hsnorm{R^TF - \id}^2}
      \quad=\quad
      \{\,\polar(F)\,\}
      \quad=\quad
      \argmin{R\,\in\,\SO(3)}{\hsnorm{F - R}^2}\;.
    \end{align*}
    This variational characterization of the orthogonal
    factor $\polar(F) \in \SO(n)$ holds in any dimension
    $n \geq 2$ (a result due to Martins and Podio-Guidugli).
    In a similar spirit, we characterize the optimal rotations
    \begin{align*}
      \rpolar_{\mu,\mu_c}(F) \;\eqdef\; \argmin{R\,\in\,\SO(n)}             {\,\left\lbrace
               \mu  \hsnorm{\sym(R^TF - \id)}^2
               \;+\;
               \mu_c\hsnorm{\skew(R^TF - \id)}^2
               \right\rbrace\,}
    \end{align*}
    for given weights $\mu > 0$ and $\mu_c \geq 0$. We identify a
    classical parameter range $\mu_c \geq \mu > 0$ for which Grioli's
    Theorem is recovered and a non-classical parameter range
    $\mu > \mu_c \geq 0$ giving rise to a new type of globally
    energy-minimizing rotations which can substantially deviate
    from $\polar(F)$. In mechanics, the weighted energy subject
    to minimization appears as the shear-stretch contribution in
    any geometrically nonlinear, quadratic, and isotropic Cosserat
    theory.
\end{minipage}
\end{center}

\vspace*{0.25cm}

\vspace*{0.125cm}
 \countres

\setcounter{tocdepth}{1}
\renewcommand{\baselinestretch}{-1.0}\normalsize
{
  \small
  \tableofcontents
}
\renewcommand{\baselinestretch}{1.0}\normalsize

\section{Introduction}\label{sec:intro}
In 1940 Giuseppe Grioli proved a variational characterization of
the orthogonal factor of the polar decomposition~\cite{Grioli40}.
In order to state this result, let $\polar(F) \in \SO(n)$ be the
unique rotation characterized as the orthogonal factor of the
right polar decomposition of
\begin{equation}
  F = \polar(F)\,U(F), \quad F \in \GL^+(n)\;,
\end{equation}
where $U(F) = \polar(F)^TF = \sqrt{F^TF} \in \Psym(n)$ denotes
the symmetric positive definite factor (which, in mechanics, is
referred to as the Biot stretch tensor).

Grioli's original result\footnote{An exposition of the original
contribution of Grioli in modernized notation has been recently
made available in~\cite{Neff_Grioli14}.} is the important special
case of space dimension $n = 3$ of the following
\begin{theo}[Grioli's theorem~\cite{Grioli40,Guidugli80,Pietraszkiewicz05}]
  \label{theo:intro:grioli}
  Let $n \geq 2$ and $\hsnorm{X}^2 \eqdef \tr{X^TX}$ the Frobenius
  norm. Then for any $F \in \GL^+(n)$, it holds
\begin{equation}
  \argmin{R\,\in\,\SO(n)}{\hsnorm{R^TF - \id}^2} \;=\; \{\polar(F)\},
  \quad\text{and thus}\quad
  \min_{R\,\in\,\SO(n)}{\hsnorm{R^TF - \id}^2}
  \;=\; \hsnorm{U - \id}^2\;.\label{intro:eq:min_energy_simple}
\end{equation}
\end{theo}
The polar factor $\polar(F) \in \SO(n)$ is the unique energy-minimizing
rotation for any given $F \in \GL^+(n)$ in any dimension $n \geq 2$,
see, e.g.,~\cite{Guidugli80}. This optimality property has an interesting
geometric interpretation following from the orthogonal invariance of the
Frobenius norm
\begin{equation}
  \hsnorm{R^TF - \id}^2 = \hsnorm{F - R}^2 = \dist^2_{\rm euclid}(F, R)
\end{equation}
which reveals a connection to the problem class of matrix distance
(or nearness) problems. In elasticity, a distance of a deformation
gradient (jacobian matrix) $F \eqdef \nabla\varphi \in \GL^+(n)$ to
a rotation $\SO(n)$ is of interest as a measure for the energy
induced by local changes in length.

In this contribution, we consider a weighted analog of Grioli's
theorem motivated by Cosserat theory and present the energy-minimizing
(optimal) rotations characterized by
\begin{prob}[Weighted optimality]
Let $n \geq 2$. Compute the set of optimal rotations
\label{intro:prob:weighted}
\begin{equation}
  \argmin{R\,\in\,\SO(n)}{\wmm(R\,;F)} \eqdef
  \argmin{R\,\in\,\SO(n)}{    \left\{
    \mu\, \hsnorm{\sym(R^TF - \id)}^2
    \,+\,
    \mu_c\,\hsnorm{\skew(R^TF - \id)}^2\right\}
  }
  \label{eq:intro:weighted}
\end{equation}
for given $F \in \GL^+(n)$ and weights $\mu > 0,\mu_c \geq 0$.
Here, $\sym(X) \eqdef \frac{1}{2}(X + X^T)$ and
$\skew(X) \eqdef \frac{1}{2}(X - X^T)$ denote the symmetric
and skew-symmetric parts of $X \in \R^{n \times n}$, respectively.
\end{prob}
Note that Grioli's theorem stated above is recovered for the
case of equal weights $\mu = \mu_c > 0$. In order to express the
connection to the variational characterization of the polar
factor $\polar(F)$, we have introduced the following notation
\begin{defi}[Relaxed polar factor(s)]
  Let $\mu > 0$ and $\mu_c \geq 0$. We denote the set-valued mapping
  that assigns to a given parameter $F \in \GL^+(n)$ its associated
  set of energy-minimizing rotations by
  \begin{equation*}
    \rpolar_{\mu,\mu_c}(F) \quad \eqdef\quad \argmin{R\,\in\,\SO(n)}{\wmm(R\,;F)}\;.
  \end{equation*}
\end{defi}
In the weighted case, the polar factor $\polar(F)$ is always critical
but \emph{not} always optimal. In general the global minimizers
$\rpolar_{\mu,\mu_c}(F)$ depend on the parameters $\mu > 0$ and
$\mu_c \geq 0$ and can substantially deviate from $\polar(F)$.

The optimal rotations in the weighted case $\rpolar_{\mu,\mu_c}(F)$
have been worked out in two and three space dimensions by the present
authors in a series of papers~\cite{Fischle:2016:OC2D,Fischle:2016:OC3D};
cf. also~\cite{Fischle:2007:PCM} and~\cite{Neff_Biot07,Neff:2009:SSNC}
for earlier related work. A visualization of the mechanism of optimal
Cosserat rotations in dimension $n = 3$ for an idealized nano-indentation
was given in~\cite{Fischle:2016:RPNI} and shows that the optimal rotations
can produce interesting non-classical patterns. A final proof of
optimality in any dimension $n \geq 2$ has been obtained by Borisov
and the authors in~\cite{Borisov:2016:ORP} and is based on a new
characterization of real square roots of real symmetric matrices.
This contribution presents an overview of these results omitting
the proofs for which we refer to the original contributions.

Our study of the energy-minimizing rotations $\rpolar_{\mu,\mu_c}(F)$
is motivated by a particular Cosserat (micropolar)
theory~\cite{Neff_Cosserat_plasticity05}, i.e., a continuum theory
with additional degrees of freedom $R \in \SO(n)$. In this
context, the objective function $\wmm(R\,;F)$ subject to
minimization in \probref{intro:prob:weighted} determines the
shear-stretch contribution to the strain energy in any
nonlinear, quadratic, and isotropic Cosserat theory, see also~\cite{Boehmer:2015:SS,Eremeyev:2012:FMM,Lankeit:2015:IC,Sansour:2008:NCC,Neff01d,Pietraszkiewicz:2009:VPN}.
The arguments to the shear-stretch energy $W_{\mu,\mu_c}(R\,;F)$ are
the deformation gradient field
$F \eqdef \nabla\varphi: \Omega \to \GL^+(n)$
and the microrotation field $R: \Omega \to \SO(n)$ evaluated at a
given point of the domain $\Omega$. A full Cosserat continuum model
furthermore contains an additional curvature energy
term~\cite{Neff_curl06} and a volumetric
energy term, see, e.g.,~\cite{Neff:2015:EGNC} or~\cite{Neff_Biot07}.

It is always possible to express the local energy contribution in a
Cosserat model as $W = W(\mstretch)$, where $\mstretch \eqdef R^TF$
is the first Cosserat deformation tensor. This reduction follows
from objectivity requirements and has already been observed by the
Cosserat brothers~\cite[p.~123, eq.~(43)]{Cosserat09},
see also~\cite{Eringen99} and~\cite{Maugin:1998:STPE}.
Since $\mstretch$ is in general non-symmetric, the most general isotropic
and quadratic local energy contribution which is zero at the reference
state is given by
\begin{equation}
  \label{intro:wgeneral}
  \underbrace{\mu\, \hsnorm{\sym(\mstretch - \id)}^2
    \,+\,
    \mu_c\,\hsnorm{\skew(\mstretch - \id)}^2}_{\text{``shear-stretch energy''}}
  \quad+\quad
  \underbrace{\frac{\lambda}{2}\,\tr{\mstretch - \id)}^2}_{\text{``volumetric energy''}}\;.
\end{equation}
The last term will be discarded in the following, since it couples
the rotational and volumetric response, a feature not present in the
well-known isotropic linear Cosserat models.\footnote{The Cosserat brothers never
  proposed any specific expression for the local energy $W = W(\mstretch)$.
  The chosen quadratic ansatz for $W = W(\mstretch)$ is motivated by a direct
  extension of the quadratic energy in the linear theory of Cosserat
  models, see, e.g.~\cite{Jeong:2009:NLIC,Neff_Jeong_bounded_stiffness09,Neff_Jeong_Conformal_ZAMM08}. We always consider a true volumetric-isochoric
  split in our applications.}

From the perspective of Cosserat theory, the optimal rotations
$\rpolar_{\mu,\mu_c}(F)$ yield insight into the important limit case
of vanishing characteristic length $L_{\rm c} = 0$.\footnote{This
  identification requires that the volume term decouples from
  the \mbox{microrotation $R$}, e.g.,
$$
W^{\rm vol}(\mstretch) \eqdef \frac{\lambda}{4} \left[\left(\det{\mstretch} - 1\right)^2 + \left(\frac{1}{\det{\mstretch}} - 1\right)^2\right]\;.
$$
This requirement is quite natural and is satisfied by all linear
Cosserat models \cite{Neff_Jeong_bounded_stiffness09,Neff_ZAMM05,Neff_Jeong_Conformal_ZAMM08}.} In this context, we can interpret the solutions
of~\eqref{eq:intro:weighted} as an energetically optimal mechanical
response of the field $R \in \SO(n)$ of Cosserat microrotations to
a given deformation gradient $F \eqdef \nabla\varphi \in \GL^+(n)$.

\begin{rem}[Vanishing Cosserat couple modulus $\mu_c$]
  The correct choice of the so-called Cosserat couple modulus
  $\mu_c \geq 0$ for specific materials and boundary value problems is
  an interesting open question. There are indications that a
  non-vanishing $\mu_c > 0$ has never been experimentally observed
  and that such a choice is at least debatable~\cite{Neff_ZAMM05}.
  The limit case $\mu_c = 0$ is hence of particular interest.
\end{rem}

We want to stress that although the term $\wmm(R\,;F)$ subject to
minimization in~\eqref{eq:intro:weighted} is quadratic in the nonsymmetric
microstrain tensor $\mstretch - \id = R^TF - \id$, see,
e.g.,~\cite{Eremeyev:2012:FMM}, the associated minimization
problem with respect to $R$ is nonlinear due to the multiplicative
coupling $R^TF$ and the geometry of $\SO(n)$.

\begin{rem}[Existence of global minimizers]
  The energy $\wmm(R\,;F)$ is a polynomial in the matrix entries,
  hence $\wmm \in C^\infty(\SO(n),\Reals)$. Further, since the Lie
  group $\SO(n)$ is compact and $\partial\!\SO(n) = \emptyset$, the
  global extrema of $\wmm$ are attained at interior points.
\end{rem}

The previous remark hints at a possible solution strategy for
\probref{intro:prob:weighted}. If \emph{all} the critical points
$R_\textrm{crit}(F) \in \SO(n)$ of $\wmm(R\,;F)$ can be
computed\footnote{The smooth manifold $\SO(n)$ has empty boundary.
This implies that a critical point for given $F \in \GL^+(n)$
satisfies $\ddtat{\wmm(R(t)\,; F)}{0} = 0$ for
every smooth curve of rotations $R(t): (-\epsilon, \epsilon) \to
\SO(n)$ passing through $R(0) = R_\textrm{crit}$.},
then a direct comparison of the associated critical energy levels
$\wmm(R_\textrm{crit}\,;F)$ allows to determine the
critical branches which are energy-minimizing.
Clearly, any minimizing critical branch realizes the
\emph{reduced} Cosserat shear-stretch energy defined as
\begin{align}
  W^{\rm red}_{\mu,\mu_c}: \GL^+(n) \to \RPosZ\;,\quad
  W^{\rm red}_{\mu,\mu_c}(F) \eqdef \min_{R\,\in\,\SO(n)} W_{\mu,\mu_c}(R\,;F)\;.
\end{align}

At first, a solution of \probref{intro:prob:weighted} in three space
dimensions was out of reach (let alone the  $n$-dimensional problem).
Therefore, we first restrict our attention to the planar case, where
we can base our computations on the standard parametrisation
\begin{equation}
  R: [-\pi, \pi] \to \SO(2) \subset \R^{2\times 2},\quad R(\malpha) \eqdef
  \begin{pmatrix}
    \cos \malpha & -\sin \malpha\\
    \sin \malpha & \cos \malpha
  \end{pmatrix}
\end{equation}
by a rotation angle.\footnote{Note that $\pi$ and $-\pi$ are mapped
to the same rotation. In this text, we implicitly choose $\pi$ over
$-\pi$ for the rotation angle whenever uniqueness is an issue.}

It turns out that there are at most two optimal planar rotations
$\rpolar_{\mu,\mu_c}^\pm(F)$ in the non-classical parameter
range $\mu > \mu_c \geq 0$ and we distinguish these by a sign.
The corresponding optimal rotation angles of
$\rpolar^\pm_{\mu,\mu_c}(F)$ are denoted by
$\alpha^\pm_{\mu,\mu_c}(F)$. The non-classical minimizers coincide
with the polar factor $\polar(F)$ in the compressive regime
of $F \in \GL^+(2)$, but deviate otherwise.

The computation of the global minimizers in dependence of $F$
is not completely obvious even for the planar case.
Hence, the following simplifications of the minimization problem
are helpful.

First, it is useful to introduce
\begin{defi}[Parameter rescaling]
  \label{defi:reduction:srad}
  \label{defi:reduction:lambda}
  \label{defi:reduction:rescaling}
  Let $\mu > \mu_c \geq 0$. We define the {\bf singular radius} $\sradmm$ by
  \begin{equation}
    \sradmm \eqdef \sradmmdef > 0\;,
    \quad\quad\text{and further define}\quad\quad
    \lambda_{\mu,\mu_c} \eqdef \frac{\sradmm}{\rho_{1,0}} = \frac{\mu}{\mu - \mu_c}\;,
  \end{equation}
  as the {\bf induced scaling parameter}. Note that $\rho_{1,0} = 2$ and
  $\lambda_{1,0} = 1$. Further, we define the {\bf parameter rescaling}
  given by
  \begin{equation}
    \widetilde{F}_{\mu,\mu_c} \;\eqdef\; \lambda^{-1}_{\mu,\mu_c}\,F \;=\; \frac{\mu - \mu_c}{\mu}\,F \quad\in \GL^+(n)\;.
  \end{equation}
\end{defi}
For $\mu > 0$ and $\mu_c = 0$, we obtain $\widetilde{F}_{\mu,0} = F$, i.e.,
the rescaling is only effective for $\mu_c > 0$.

Regarding the material parameters, we proved
in~\cite{Fischle:2016:OC2D} that for any dimension
$n \geq 2$, it is in fact sufficient to restrict
our attention to two parameter pairs:
$(\mu,\mu_c) = (1,1)$, the \emph{classical} case, and
$(\mu,\mu_c) = (1,0)$, the \emph{non-classical} case.
Hence, somewhat surprisingly, the solutions for arbitrary
$\mu > 0$ and $\mu_c \geq 0$ can be recovered from these
two limit cases. This is the content of
\begin{lem}[Parameter reduction]
  \label{lem:parameter_reduction}
  Let $n \geq 2$ and let $F \in \GL^+(n)$, then
  \begin{equation}
    \begin{aligned}
      \mu_c \geq \mu > 0 \quad&\Longrightarrow\quad \wmm(R\,;F)
      \;\sim\;
      W_{1,1}(R\,;F)\;,\quad\text{and}\\
      \mu > \mu_c \geq 0 \quad&\Longrightarrow\quad \wmm(R\,;F)
      \;\sim\;
      \wsym(R\,;\widetilde{F}_{\mu,\mu_c})\;.
    \end{aligned}
  \end{equation}
\end{lem}
Here, the equivalence notation means that the energies give rise to
the same global minimizers which we can also state as
\begin{cor}
  \begin{equation}
    \rpolar_{\mu,\mu_c}(F) =
    \begin{cases}
      \rpolar_{1,1}(F) = \{\polar(F)\}, & \text{if}\quad \mu_c \geq \mu > 0\\
      \rpolar_{1,0}(\widetilde{F}_{\mu,\mu_c}), & \text{if}\quad \mu > \mu_c \geq 0
    \end{cases}
  \end{equation}
\end{cor}

Another important observation can be made introducing the rotation
\begin{equation}
  \label{eq:Rhat}
  \widehat{R} \eqdef Q^TR^T\polar Q
\end{equation}
which acts \emph{relative} to the polar factor $\polar(F)$ in the
coordinate system given by the columns of $Q$ which span a positively
oriented frame of principal directions of $U$. This allows us to
transform
\begin{align}
  &Q^T(\sym(R^TF) - \id)Q = Q^T\left(\sym(R^T\polar QDQ^T) - \id)\right)Q\notag\\
  &\quad=\sym(Q^TR^T\polar QDQ^TQ - Q^TQ) = \sym(\underbrace{Q^TR^T\polar Q}_{\defeq\;\widehat{R}}D - \id) = \sym(\widehat{R}D - \id)\;.\label{eq:symtransform}
\end{align}
For fixed choice of $Q \in \SO(n)$, the inverse transformation allows
to reconstruct the absolute rotation uniquely
\begin{equation}
  \label{eq:R}
  R = \left(Q\widehat{R}Q^T\polar ^T\right)^T = \polar Q\widehat{R}^TQ^T\;.
\end{equation}
Hence, in the non-classical parameter range represented by the
limit case $(\mu,\mu_c) = (1,0)$, the minimization problem can
be reduced to the following problem for the optimal relative
rotations.
\begin{prob}
  \label{prob:opt}
  Let $n \geq 2$. Compute the set of energy-minimizing relative
  rotations
  \begin{equation}
    \rpolar_{1,0}(D) \;\eqdef\;
    \argmin{\widehat{R}\,\in\,\SO(n)}{W_{1,0}(\widehat{R}\,;D)}
    \;=\;
    \argmin{\widehat{R}\,\in\,\SO(n)}{\hsnorm{\sym(\widehat{R}D - \id)}^2} \;\subseteq\; \SO(n)\;.
  \end{equation}
\end{prob}
The decisive point in the solution of \probref{prob:opt}
in dimensions $n \geq 3$ is the characterization of the set
of relative rotations $\widehat{R} \in \SO(n)$ satisfying the
particular symmetric square condition
$$(\widehat{R}D - \id)^2 \in \Sym(n)$$
which is equivalent to the Euler-Lagrange equations.

After having set the stage of the optimization problem on $\SO(n)$,
this overview is now structured as follows: in the next Section 2,
we consider in some detail the planar problem which allows for a
complete solution by elementary techniques and which presents
already the essential geometry which unfolds in dimensions
$n \geq 3$. In Section 3, we provide the complete solution for
the three-dimensional case as well as the corresponding reduced
energy expression in terms of singular values of $F$. We also
provide a geometrical interpretation that allows to view the
minimization problem for $\mu_c = 0$ as a distance problem.
Furtermore, we provide a discussion for which deformation
gradients we can only have the classical response $\polar(F)$.
Finally, in Section 4, we present our results for the general
$n$-dimensional case.

\countres
\section{Optimal rotations in two space dimensions}

In this section, we consider
\begin{prob}[The planar minimization problem]
  \label{intro:prob:planar}
  Let $F \in \GL^+(2)$, $\mu > 0$ and $\mu_c \geq 0$. The task is to
  compute the set of optimal microrotation angles
\begin{equation}
  \argmin{\malpha\;\in\;[-\pi,\pi]}{
    \left\lbrace
    \mu \hsnorm{\sym(R(\alpha)^TF - \id_2)}^2
    +
    \mu_c \hsnorm{\sym(R(\alpha)^TF - \id_2)}^2
    \right\rbrace
  }\;,
  \end{equation}
  where
  \begin{equation*}
    R(\alpha) \eqdef
    \begin{pmatrix}
      \cos \malpha & -\sin \malpha\\
      \sin \malpha & \cos \malpha
    \end{pmatrix} \in \SO(2)
    \quad \text{and} \quad
    \begin{pmatrix}
    F_{11} & F_{12}\\
    F_{21} & F_{22}\\
    \end{pmatrix} \in \GL^+(2)\;.
    \end{equation*}
\end{prob}

In this case we can compute explicit representations of optimal
planar rotations for the Cosserat shear-stretch energy by
elementary means. The parameter reduction strategy described
by \leref{lem:parameter_reduction} allows us to concentrate
our efforts towards the construction of explicit solutions
to~\probref{intro:prob:planar} on two representative pairs of parameter
values $\mu$ and $\mu_c$. The classical
regime is characterized by the limit case $(\mu,\mu_c) = (1,1)$ and
the unique minimizer is given by the polar factor $\polar(F)$ for
any dimension $n \geq 2$.

The non-classical case represented by $(\mu,\mu_c) = (1,0)$ turns
out to be much more interesting and we compute all global non-classical
minimizers $\rpolar_{1,0}(F)$ for $n = 2$. This is the main contribution
of this section. Furthermore, we derive the associated
reduced energy levels $W^{\rm red}_{1,1}(F)$ and $W^{\rm red}_{1,0}(F)$
which are realized by the corresponding optimal Cosserat microrotations.
Finally, we reconstruct the minimizing rotation angles for general values
of $\mu$ and $\mu_c$ from the classical and non-classical limit cases.

\subsection{Explicit solution for the classical parameter range: $\mu_c \geq \mu > 0$}
The polar factor $\polar(F)$ is uniquely optimal for the classical
parameter range in any dimension $n \geq 2$. Let us give an explicit
representation for $n = 2$ in terms of
$\palpha \in (-\pi,\pi]$. In view of the parameter reduction, distilled
in~\leref{lem:parameter_reduction}, it suffices to compute the set of
optimal rotation angles for the representative limit case
$(\mu,\mu_c) = (1,1)$.

Thus, to obtain an explicit representation of $\palpha \in (-\pi,\pi]$ which
characterizes the polar factor $\polar(F)$ in dimension $n = 2$, we
consider
\begin{equation}
  \argmin{\malpha\;\in\;[-\pi,\pi]} W_{1,1}(R(\malpha)\,;F)
  = \argmin{\malpha\;\in\;[-\pi,\pi]}{\hsnorm{\left[\begin{pmatrix}
      \cos \malpha & -\sin \malpha\\
      \sin \malpha & \cos \malpha
\end{pmatrix}^T
\begin{pmatrix}
    F_{11} & F_{12}\\
    F_{21} & F_{22}\\
\end{pmatrix}
-\begin{pmatrix}
 1 & 0\\
 0 & 1
\end{pmatrix}\right]}^2}\,.
\end{equation}
Let us introduce the rotation
$J \eqdef \begin{pmatrix} 0 & -1\\ 1 & 0 \end{pmatrix} \in \SO(2)$.
Its application to a vector $v \in \R^2$ corresponds to multiplication
with the imaginary unit $i \in \C$. In what follows, the quantities
$\tr{F} = F_{11} + F_{22}$ and $\tr{JF} = - F_{21} + F_{12}$
play a particular role and we note the identity
\begin{equation}
\tr{F}^2 + \tr{JF}^2 = \hsnorm{F}^2 + 2\,\det{F} = \tr{U}^2\;.
\end{equation}

The reduced energy
$W_{1,1}^{\rm red}(F) \eqdef \min_{R\in\SO(n)} W_{1,1}(R\,;F)$
realized by the polar factor $\polar(F)$ can be shown to be the euclidean
distance of an arbitrary $F$ in $\R^{n\times n}$ to $\SO(n)$. For $n = 2$,
we obtain
\begin{theo}[Euclidean distance to planar rotations]
  Let $F \in \GL^+(2)$, then
  \begin{equation}
    W_{1,1}^{\rm red}(F) = \dist^2(F,\SO(2)) = \hsnorm{U - \id}^2
    = \hsnorm{F}^2 - 2\,\sqrt{\hsnorm{F}^2 + 2\,\det{F}} + 2\;.
  \end{equation}
  The unique optimal rotation angle realizing this minimial
  energy level satisfies the equation
  \begin{equation}
    \begin{pmatrix}
      \sin\palpha\\
      \cos\palpha
    \end{pmatrix}
    =\frac{1}{\tr{U}}
    \begin{pmatrix}
      -\tr{JF} \\
      \tr{F}
    \end{pmatrix}\;.
  \end{equation}
  In particular, we have
  $\palpha(F) = -\sign(\tr{JF})\cdot\arccos\left(\frac{\tr{F}}{\tr{U}}\right) \in [-\pi,\pi]$.
\end{theo}

\begin{cor}[Explicit formula for $\polar(F)$]
  \label{cor:polar_planar}
  Let $F \in \GL^+(2)$, then the polar factor $\polar(F)$ has the
  explicit representation
  \begin{equation}
    \polar(F) =
    R(\palpha) \eqdef \begin{pmatrix}
      \cos\palpha & -\sin\palpha\\
      \sin\palpha &  \cos\palpha
    \end{pmatrix}
    = \frac{1}{\tr{U}}
    \begin{pmatrix}
      \phantom{-}\tr{F}  & \tr{JF}\\
      -\tr{JF} & \tr{F}
    \end{pmatrix}\;.
  \end{equation}
\end{cor}

\subsection{The limit case $(\mu,\mu_c) = (1,0)$ for $\mu > \mu_c \geq 0$}
We now approach the more interesting non-classical limit case
$(\mu, \mu_c) = (1,0)$ and compute the optimal rotations for
$W_{\mu,\mu_c}(R\,;F)$. Note that, due to~\leref{lem:parameter_reduction},
this limit case represents the entire non-classical parameter
range $\mu > \mu_c \geq 0$.

\begin{theo}[The formally reduced energy $W^{\rm red}_{1,0}(F)$]
  \label{theo:wred10}
  Let $F \in \GL^+(2)$. Then, the {\bf formally reduced energy}
  \begin{equation}
    W_{1,0}^{\rm red}(F) \eqdef \min_{R \; \in \; \SO(2)}W_{1,0}(R\,;F)
                       \eqdef \min_{R\,\in\,\SO(2)} \norm{\sym(R^TF-\id)}^2
  \end{equation}
is given by
\begin{equation}
  \label{eq:minimization:wsymred}
    W_{1,0}^{\rm red}(F)
    = \begin{cases}
      \hsnorm{U - \id}^2 = \tr{(U - \id)^2} = \dist^2(F,\SO(2))\;,   &\quad\text{if} \quad \tr{U} < 2\\
      \frac{1}{2}\hsnorm{F}^2 - \det{F}
                = \frac{1}{2}\,\tr{U}^{2} - 2\,\det{U}\;, &\quad\text{if}\quad \tr{U} \ge 2\;.
    \end{cases}
  \end{equation}
\end{theo}

It is well-known that any orthogonally invariant energy density
$W(F)$ admits a representation in terms of the singular values of $F$, i.e.,
in the eigenvalues of $U$. Let us give this representation.
\begin{cor}[Representation of $W^{\rm red}_{1,0}(F)$ in the singular values
    of $F$]
  \label{cor:wred10_sing}
  Let $F \in \GL^+(2)$ and denote its singular values by $\nu_i$, $i = 1,2$.
  The representation of $W^{\rm red}_{1,0}(F)$ in the singular values of $F$
  is given by
  \begin{equation}
    W_{1,0}^{\rm red}(F) = W_{1,0}^{\rm red}(\nu_1,\nu_2) =
    \begin{cases}
      (\nu_1 - 1)^2 + (\nu_2 - 1)^2\;,  &\text{if}\quad \nu_1 + \nu_2  < 2\\
      \frac{1}{2}(\nu_1 - \nu_2)^2\;,  &\text{if}\quad\nu_1 + \nu_2 \geq 2\;.
    \end{cases}
  \end{equation}\
\end{cor}
Note that the previous formulae are independent of the enumeration
of the singular values.

\subsubsection{Optimal relative rotations for \texorpdfstring{$\mu = 1$}{mu = 1} and \texorpdfstring{$\mu_c = 0$}{muc = 0}}

Our next goal is to compute explicit representations of the rotations
$\rpolar^\pm_{1,0}(F)$ which realize the minimal energy level in the
non-classical limit case $(\mu,\mu_c) = (1,0)$. This is the content
of the next theorem for which we now prepare the stage with the
following
\begin{lem}
  \label{lem:eq_tr_hatrd_eq_2}
  Let $D = \diag(\sigma_1,\sigma_2) > 0$, i.e, a diagonal matrix with
  strictly positive diagonal entries. Then, assuming $\tr{D} \geq 2$,
  the equation $\tr{R(\beta)\,D} = 2$ has the following solutions
    \begin{equation}
      \beta^\pm
      \;=\;
      \pm\,\arccos\left(\frac{2}{\tr{D}}\right) \quad\in [-\pi,\pi]\;.
    \end{equation}
    For $\tr{D} < 2$, there exists no solution, but we can define
    $\beta = \beta^\pm \eqdef 0$ by continuous extension.
\end{lem}

\begin{figure}
  \begin{center}
    \includegraphics{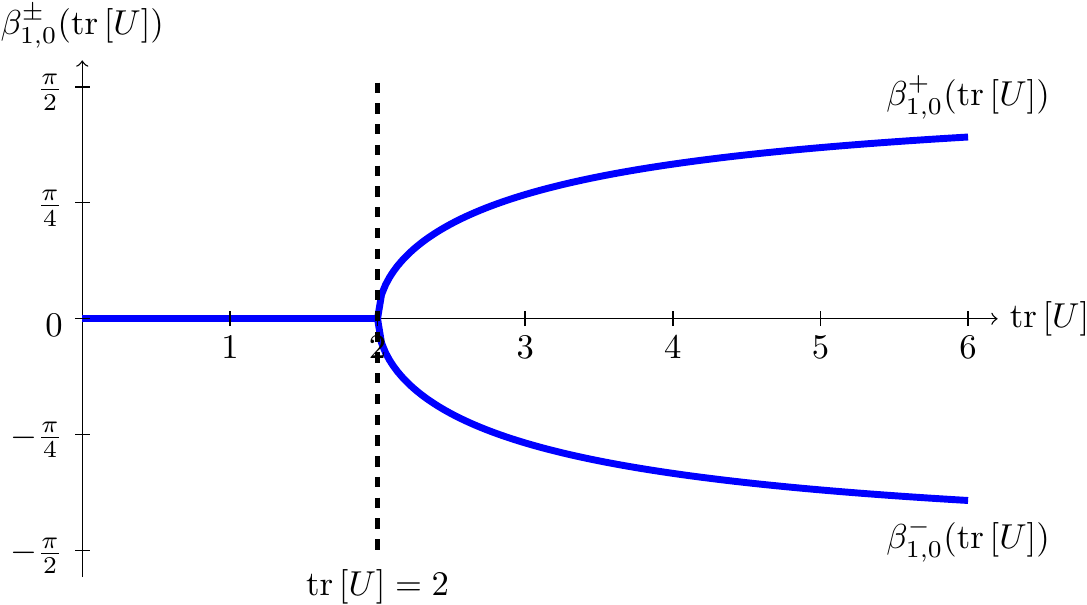}
      \end{center}
  \caption[The optimal relative rotation angle $\beta_{1,0}$ as a function of $\tr{U}$.]{
    Plot of the two optimal relative rotation angles
    $\beta_{1,0}^\pm = \pm\arccos\left(\frac{2}{\tr{U}}\right)$
    for the non-classical limit case $(\mu,\mu_c) = (1,0)$.
    Note the pitchfork bifurcation in $\tr{U} = \rho_{1,0} = 2$.
    For $0 < \tr{U} < 2$, the polar angle $\palpha$ is uniquely
    optimal and the relative rotation angle $\beta$ vanishes
    identically.\label{fig:alpha_rel_plot}    }
\end{figure}

Our~\figref{fig:alpha_rel_plot} shows a plot of the optimal
relative rotation angle $\beta(\tr{U})$.
In the classical parameter range $0 < \tr{U} \leq 2$,
$\palpha(F)$ is uniquely optimal and $\beta$ vanishes identically.
In \mbox{$\tr{U} = 2$}, a classical pitchfork bifurcation occurs. In
particular, due to $\tr{U(\id_2)} = \tr{\id_2} = 2$,
the identity matrix is a bifurcation point of $\beta^\pm(F)$.
Further, we note that the branches $\beta^\pm(\tr{U}) = \pm\arccos(2/\tr{U})$
\emph{are not differentiable} at $\tr{U} = 2$. This has implications
on the interaction of the Cosserat shear-stretch energy with the
Cosserat curvature energy $W_{\rm curv}$.

\begin{theo}[Optimal non-classical microrotation angles $\malpha_{1,0}^\pm$]
  \label{theo:malpha10pm}
  Let $F \in \GL^+(2)$ and consider $(\mu,\mu_c) = (1,0)$. The optimal
  rotation angles for $\wsym$ are given by
  \begin{equation}
    \malpha_{1,0}^{\pm}(F) =
    \begin{cases}
      \palpha(F) = \arccos(\frac{\tr{F}}{\tr{U}})
      & ,\quad\text{if}\qquad \tr{U} < 2\\
      \palpha(F) \pm\,\arccos\left(\frac{2}{\tr{U}}\right)
      & ,\quad\text{if}\qquad \tr{U} \geq 2\;.
    \end{cases}
  \end{equation}
\end{theo}

\subsection{Expressions for general non-classical parameter choices}
\label{subsec:minimization:reconstruction}
The reduction for $\mu$ and $\mu_c$ in~\leref{lem:parameter_reduction}
asserts that the optimal rotations for arbitrary values of
$\mu > 0$ and $\mu_c \geq 0$
can be reconstructed from the limit cases $(\mu, \mu_c) = (1,1)$ and
$(\mu, \mu_c) = (1,0)$. We now detail this procedure which essentially
exploits~\deref{defi:reduction:rescaling}.

Note first that the rescaled deformation gradient $\widetilde{F}_{\mu,\mu_c} \eqdef \lambda^{-1}_{\mu,\mu_c} F$ induces a rescaled stretch tensor
\begin{equation}
  \widetilde{U}_{\mu,\mu_c} = \sqrt{(\widetilde{F}_{\mu,\mu_c})^T\widetilde{F}_{\mu,\mu_c}} = \lambda^{-1}_{\mu,\mu_c}\cdot U\;.
\end{equation}
The right polar decomposition takes the form
$\widetilde{F}_{\mu,\mu_c} = \polar(\widetilde{F}_{\mu,\mu_c})\,\widetilde{U}_{\mu,\mu_c}$.
From $\polar(\widetilde{F}_{\mu,\mu_c}) = \widetilde{F}_{\mu,\mu_c}\widetilde{U}_{\mu,\mu_c}^{-1}$ follows the scaling invariance $\polar(\widetilde{F}_{\mu,\mu_c}) = \polar(F)$.
For the non-classical parameter range $\mu > \mu_c \geq 0$, the quantity
\begin{equation}
  \tr{\widetilde{U}_{\mu,\mu_c}} = \tr{ \lambda_{\mu,\mu_c}^{-1} \cdot U} = \frac{\rho_{1,0}}{\rho_{\mu_,\mu_c}}\tr{U}
\end{equation}
plays an essential role. This leads us to
\begin{equation}
  \tr{\widetilde{U}_{\mu,\mu_c}} \geq 2 = \rho_{1,0}
  \quad \isequivto \quad
  \tr{\frac{\rho_{1,0}}{\rho_{\mu_,\mu_c}}\cdot U} \geq \rho_{1,0}
  \quad \isequivto \quad
  \tr{U} \geq \rho_{\mu,\mu_c}\;.
\end{equation}
In particular, this implies that the bifurcation in $\tr{U}$
allowing for non-classical optimal planar rotations is
characterized by the singular radius $\sradmm \eqdef \sradmmdef$.

\begin{theo}
  Let $F \in \GL^+(2)$. For $\mu_c \geq \mu > 0$
  the optimal microrotation angle is given by
  \begin{equation}
    \alpha_{\mu,\mu_c}(F)
    = \palpha(\widetilde{F}_{\mu,\mu_c})
    = \palpha(F) = \arccos\left(\frac{\tr{F}}{\tr{U}}\right)\;.
  \end{equation}
  For $\mu > \mu_c \geq 0$, the two optimal rotation angles
  are given by
  \begin{equation}
    \malpha^{\pm}_{\mu,\mu_c}(F) = \malpha^{\pm}_{1,0}(\widetilde{F}_{\mu,\mu_c})
  = \begin{cases}
    \palpha(F) = \arccos\left(\frac{\tr{F}}{\tr{U}}\right) &,\quad\text{if}\qquad \tr{U} < \rho_{\mu,\mu_c}\\
    \palpha(F) \mp\,\arccos\left(\frac{\sradmm}{\tr{U}}\right)&,\quad\text{if}\qquad \tr{U} \geq \rho_{\mu,\mu_c}\;.
  \end{cases}
\end{equation}
\end{theo}

 \countres
\makeatletter{}\subsection{Optimal rotations for planar simple shear}
\label{sec:shear}
We now apply our previous optimality results to simple shear
deformations
\begin{equation}
  F_{\gamma} \eqdef
  \begin{pmatrix}
    1 & \gamma\\
    0 & 1
  \end{pmatrix}, \quad\gamma \in \R\;.
\end{equation}
The energy-minimizing rotation angles
$\malpha_{\mu,\mu_c}(\gamma) \eqdef \malpha_{\mu,\mu_c}(F_\gamma)$
for simple shear can be explicitly computed; see also~\cite{Neff:2009:SSNC}
for previous results.

In the classical parameter range $\mu_c \geq \mu > 0$ represented
by the limit case $(\mu,\mu_c) = (1,1)$ the polar rotation
$\polar(F_\gamma)$ is uniquely optimal.

Let us collect some properties of simple shear $F_\gamma$. We have
$\hsnorm{F_\gamma}^2 = 2 + \gamma^2$ and $\det{F_{\gamma}} = 1$, i.e.,
simple shear is volume preserving for any amount $\gamma$. This
allows us to compute
\begin{equation}
  \tr{U_\gamma} \;=\; \sqrt{\hsnorm{F_\gamma}^2 + 2\,\det{F_\gamma}}
               \;=\; \sqrt{4 + \gamma^2} \geq 2\;.
\end{equation}
Thus, we have
\begin{cor}[Optimal non-classical Cosserat rotations for simple shear]
  Let $(\mu,\mu_c) = (1,0)$ and let $F_{\gamma} \in \GL^+(2)$ be a simple
  shear of amount $\gamma \in \R$. Then,
  \begin{equation} \gamma \neq 0
    \quad\Longrightarrow\quad
    \rpolar^\pm_{1,0}(F_\gamma) \;\mathbf{\neq}\; \polar(F_\gamma)\;.
  \end{equation}
\end{cor}

\begin{rem}[Symmetry of the first Cosserat deformation
            tensor $\mstretch$ in simple shear]
  A simple shear $F_\gamma$ by a non-zero amount $\gamma \neq 0$
  automatically generates an optimal microrotational response
  $\rpolar^\pm(F_\gamma)$ which deviates from the continuum
  rotation $\polar(F)$. This implies that the associated first
  Cosserat deformation tensor $\mstretch^\pm_{1,0}(F_\gamma) \eqdef \rpolar_{1,0}^\pm(F_\gamma)^TF_\gamma$ is not symmetric for any
  $\gamma \neq 0$.
\end{rem}

\countres
\section{Optimal rotations in three space dimensions}
In this section, we discuss
\begin{prob}[Weighted optimality in dimension $n = 3$]
  Let $\mu > 0$ and $\mu_c \geq 0$. Compute the set of optimal rotations
  \label{intro:prob_wmm}
  \begin{equation}
    \argmin{R\,\in\,\SO(3)}{\wmm(R\,;F)}
    \;\eqdef\;
    \argmin{R\,\in\,\SO(3)}{      \left\{
      \mu\, \hsnorm{\sym(R^TF - \id)}^2
      \,+\,
      \mu_c\,\hsnorm{\skew(R^TF - \id)}^2\right\}
    }
    \label{eq:intro:weighted_wmm}
  \end{equation}
  for given parameter $F \in \GL^+(3)$ with distinct singular values
  $\nu_1 > \nu_2 > \nu_3 > 0$.
\end{prob}

The polar factor $\polar(F)$ is the \emph{unique} minimizer for
$W_{\mu,\mu_c}(R\,;F)$ in the \emph{classical} parameter range
$\mu_c \geq \mu > 0$, in all dimensions $n \geq 2$,
see~\cite{Guidugli:1980:EPP,Neff_Grioli14}.

Since the classical parameter domain $\mu_c \geq \mu > 0$ is very well
understood, we focus entirely on the non-classical parameter
range $\mu > \mu_c \geq 0$. Furthermore, due to the parameter
reduction described by \leref{lem:parameter_reduction}, which holds
for all dimensions $n \geq 2$, it suffices to solve the non-classical
limit case $(\mu,\mu_c) = (1,0)$, since
\begin{equation}
  \argmin{R\,\in\,\SO(3)}{W_{\mu,\mu_c}(R\,;F)}
  \quad=\quad
  \argmin{R\,\in\,\SO(3)}{W_{1,0}(R\,; \widetilde{F}_{\mu,\mu_c})}\;.
\end{equation}
On the right hand side, we notice a \emph{rescaled deformation gradient}
$$\widetilde{F}_{\mu,\mu_c} \eqdef \lambda^{-1}_{\mu,\mu_c} \cdot F \in \GL^+(3)$$
which is obtained from $F \in \GL^+(3)$ by multiplication with the inverse of
the \emph{induced scaling parameter}
$\lambda_{\mu,\mu_c} \eqdef \frac{\mu}{\mu - \mu_c} > 0$. We note that we use
the previous notation throughout the text and further introduce
the \emph{singular radius} $\rho_{\mu,\mu_c} \eqdef \frac{2\mu}{\mu - \mu_c}$.

It follows that the set of optimal Cosserat rotations can be described
by
\begin{equation}
  \rpolar_{\mu,\mu_c}(F) \;=\; \rpolar_{1,0}(\widetilde{F}_{\mu,\mu_c})
\end{equation}
for the entire non-classical parameter range $\mu > \mu_c \geq 0$.
We are therefore mostly concerned with the case $\mu_c = 0$ in the
present text. Note that for all $\mu > 0$, we have the equality
\begin{equation}
  \rpolar^\pm_{\mu,0}(F)
  \;=\;
  \rpolar^\pm_{1,0}(F)\;.
\end{equation}

\subsection{The locally energy-minimizing Cosserat rotations $\rpolar^\pm_{\mu,\mu_c}(F)$}
We briefly present the geometric characterization of the optimal
Cosserat rotations $\rpolar^\pm_{\mu,\mu_c}(F)$ obtained
in~\cite{Fischle:2016:OC3D}. Let $R \in \SO(3)$ and
let $\S^2 \subset \R^3$ denote the unit $2$-sphere.
We make use of the well-known angle-axis parametrization of
rotations which we write as $[\alpha,\, r]$\footnote{The angle-axis parametrization
  is singular, but this is not an issue for our exposition.},
where $\alpha \in (-\pi,\pi]$ denotes the rotation angle
and $r \in \S^{2}$ specifies the oriented rotation axis.

We recall that it is sufficient to solve for the relative
rotation, i.e., we consider
\begin{prob}[Diagonal form of weighted optimality in $n = 3$]
  \label{intro:prob_wmm_reduced}
  Let $\mu > 0$ and $\mu_c \geq 0$ and let
  $D = \diag(\nu_1,\nu_2,\nu_3)$ with
  $\nu_1 > \nu_2 > \nu_3 > 0$. Compute the set of
  optimal relative rotations
  \begin{equation}
    \argmin{\widehat{R}\,\in\,\SO(3)}{\wmm(\widehat{R}^T\,;D)}
    \;\eqdef\;
    \argmin{\widehat{R}\,\in\,\SO(3)}{      \left\{
      \mu\, \hsnorm{\sym(\widehat{R}\,D - \id)}^2
      \,+\,
      \mu_c\,\hsnorm{\skew(\widehat{R}\,D - \id)}^2\right\}
    }\;.
  \end{equation}
\end{prob}
We stress that the rotation angle of the
relative rotation $\widehat{R}$ is implicitly reversed
due to the correspondence $R^T \leftrightarrow \widehat{R}$.

The computation of the solutions to~\probref{intro:prob_wmm_reduced}
by computer algebra together with a statistical verification are
the core results obtained in~\cite{Fischle:2016:OC3D} which we
present next.

\begin{propo}[Energy-minimizing relative rotations
              for $(\mu,\mu_c) = (1,0)$]
  \label{propo:rhat}
  Let $\nu_1 > \nu_2 > \nu_3 > 0$ be the singular values of
  $F \in \GL^+(3)$. Then the energy-minimizing relative rotations
  solving~\probref{intro:prob_wmm_reduced} are given by
  \begin{equation}
    \widehat{R}_{1,0}^{\pm}(F)
    \quad\eqdef\quad
    \begin{pmatrix}
      \cos \hat{\beta}^\pm_{1,0}  & -\sin \hat{\beta}^\pm_{1,0} & 0\\
      \sin \hat{\beta}^\pm_{1,0}  &  \cos \hat{\beta}^\pm_{1,0} & 0\\
      0                    &  0                   & 1\\
    \end{pmatrix}\;,
  \end{equation}
  where the optimal rotation angles
  $\hat{\beta}^\pm_{1,0} \in (-\pi,\pi]$
  are given by
    \begin{equation}
      \hat{\beta}^\pm_{1,0}(F)
      \quad\eqdef\quad
      \begin{cases}
        \; 0\;,  &\quad\quad\text{if}\quad \nu_1 + \nu_2 \leq 2\;,\\
        \; \pm\arccos(\frac{2}{\nu_1 + \nu_2})\;,
        &\quad\quad\text{if}\quad \nu_1 + \nu_2 \geq 2\;.
      \end{cases}
    \end{equation}
    Thus, in the non-classical regime $\nu_1 + \nu_2 \geq 2$,
    we obtain the explicit expression
    \begin{equation}
      \widehat{R}_{1,0}^{\pm}(F)
      \quad\eqdef\quad
      \begin{pmatrix}
        \frac{2}{\nu_1 + \nu_2}  & \mp \sqrt{1-\left(\frac{2}{\nu_1 + \nu_2}\right)^2} & 0\\
        \pm \sqrt{1-\left(\frac{2}{\nu_1 + \nu_2}\right)^2} & \frac{2}{\nu_1 + \nu_2} & 0 \\
        0 & 0 & 1
      \end{pmatrix}\;.
    \end{equation}
    In the classical regime $\nu_1 + \nu_2 \leq 2$, we simply
    obtain the relative rotation $\widehat{R}_{1,0}^{\pm}(F) = \id$,
    and there is no deviation from the polar factor $\polar(F)$
    at all.
\end{propo}

Note that, due to the parameter reduction~\leref{lem:parameter_reduction},
it is always possible to recover the optimal rotations
$\rpolar^\pm_{\mu,\mu_c}(F)$
for general non-classical parameter choices $\mu > \mu_c \geq 0$
from the non-classical limit case
$(\mu,\mu_c) = (1,0)$; cf.~\cite{Fischle:2016:OC2D}
and~\cite{Fischle:2016:OC3D} for details.

\subsection{Geometric and mechanical aspects of optimal Cosserat
            rotations}

It seems natural to introduce
\begin{defi}[Maximal mean planar stretch and strain]
  \label{defi:mmpss}
  Let $F \in \GL^+(3)$ with singular values
  $\nu_1 \geq \nu_2 \geq \nu_3 > 0$. We introduce
  the \textbf{maximal mean planar stretch} $\ump$ and
  the \textbf{maximal mean planar strain} $\smp$ as follows:
  \begin{equation}
    \begin{aligned}
      \ump(F) &\;\eqdef\; \frac{\nu_1 + \nu_2}{2}\;,\quad\text{and}\\
      \smp(F) &\;\eqdef\; \frac{(\nu_1 - 1) + (\nu_2 - 1)}{2} = \ump(F) - 1\;.
    \end{aligned}
  \end{equation}
\end{defi}

In order to describe the bifurcation behavior of $\rpolar_{\mu,\mu_c}^\pm(F)$
as a function of the parameter $F \in \GL^+(3)$, it is helpful to
partition the parameter space $\GL^+(3)$.
\begin{defi}[Classical and non-classical domain]
To any pair of material parameters $(\mu,\mu_c)$ in the non-classical
range $\mu > \mu_c \geq 0$, we associate a \textbf{classical domain}
$\domc_{\mu,\mu_c}$ and a \textbf{non-classical domain} $\domn_{\mu,\mu_c}$.
Here,
\begin{equation}
  \begin{aligned}
    \domc_{\mu,\mu_c} &\eqdef \setdef{F \in \GL^+(3)}{\smp(\widetilde{F}_{\mu,\mu_c}) \leq 0}\;,
    \quad\text{and}\quad\\
    \domn_{\mu,\mu_c} &\eqdef \setdef{F \in \GL^+(3)}{\smp(\widetilde{F}_{\mu,\mu_c}) \geq 0}\;,
  \end{aligned}
\end{equation}
respectively.
\end{defi}

It is straight-forward to derive the following equivalent characterizations
{\small
\begin{equation}
  \begin{aligned}
    \domc_{\mu,\mu_c} = \setdef{F \in \GL^+(3)}{\ump(F) \leq \lambda_{\mu,\mu_c}}
    = \setdef{F \in \GL^+(3)}{\nu_1 + \nu_2 \leq \sradmm \eqdef \frac{2\mu}{\mu - \mu_c}}\;,\\
    \domn_{\mu,\mu_c} = \setdef{F \in \GL^+(3)}{\ump(F) \geq \lambda_{\mu,\mu_c}}
    = \setdef{F \in \GL^+(3)}{\nu_1 + \nu_2 \geq \sradmm \eqdef \frac{2\mu}{\mu - \mu_c}}\;.
  \end{aligned}
\end{equation}
}
On the intersection
$\domc_{\mu,\mu_c} \cap \domn_{\mu,\mu_c} = \setdef{F \in \GL^+(3)}{\smp(F) = 0}$,
the minimizers $\rpolar_{\mu,\mu_c}^\pm(F)$ coincide with the polar
factor $\polar(F)$. This can be seen from the form of the optimal
relative rotations in~\proporef{propo:rhat}. More explicitly, in
dimension $n = 3$ and in the non-classical limit case
$(\mu,\mu_c) = (1,0)$, we have:
\begin{equation}
  \domc_{1,0} \eqdef \setdef{F \in \GL^+(3)}{\smp(F) \leq 0}\;,
  \quad\text{and}\quad
  \domn_{1,0} \eqdef \setdef{F \in \GL^+(3)}{\smp(F) \geq 0}\;.
\end{equation}
Since the maximal mean planar strain $\smp(F)$ is related to
strain, this indicates a particular (possibly new) type of
tension-compression asymmetry.

Towards a geometric interpretation of the energy-minimizing Cosserat
rotations $\rpolar^\pm_{1,0}(F)$ in the non-classical limit case
$(\mu,\mu_c) = (1,0)$, we reconsider the spectral decomposition of
$U = QDQ^T$ from the principal axis transformation in~\secref{sec:intro}.
Let us denote the columns of $Q \in \SO(3)$ by $q_i \in \S^2$, $i = 1,2,3$.
Then $q_1$ and $q_2$ are orthonormal eigenvectors of $U$ which correspond
to the largest two singular values $\nu_1$ and $\nu_2$ of $F \in \GL^+(3)$.
More generally, we introduce the following

\begin{defi}[Plane of maximal stretch]
  \label{defi:pms}
  The \textbf{plane of maximal stretch} is the linear subspace
  $$\mathrm{P}^{\rm ms}(F) \quad\eqdef\quad \vspan{q_1,q_2} \subset \R^3$$
  spanned by the two eigenvectors $q_1,q_2$ of $U$ associated with the two
  largest singular values $\nu_1 > \nu_2 > \nu_3 > 0$
  of the deformation gradient $F \in \GL^+(3)$.
\end{defi}

We recall that, due to the parameter
reduction~\leref{lem:parameter_reduction}, it is always possible to
recover the optimal rotations
\begin{equation}
  \rpolar_{\mu,\mu_c}(F) \eqdef \argmin{R\,\in\,\SO(3)}{W_{\mu,\mu_c}(R\,;F)}
\end{equation}
for a general choice of non-classical parameters $\mu > \mu_c \geq 0$ from the
non-classical limit case $(\mu,\mu_c) = (1,0)$. However, we defer
the explicit procedure for a bit since it is quite instructive
to interpret this distinguished non-classical limit case first.
\begin{center}
  \begin{tikzpicture}
    \node at (0,0){\includegraphics[width=.41\linewidth,clip=true,trim=0.75cm 3cm 0.75cm 2.5cm]{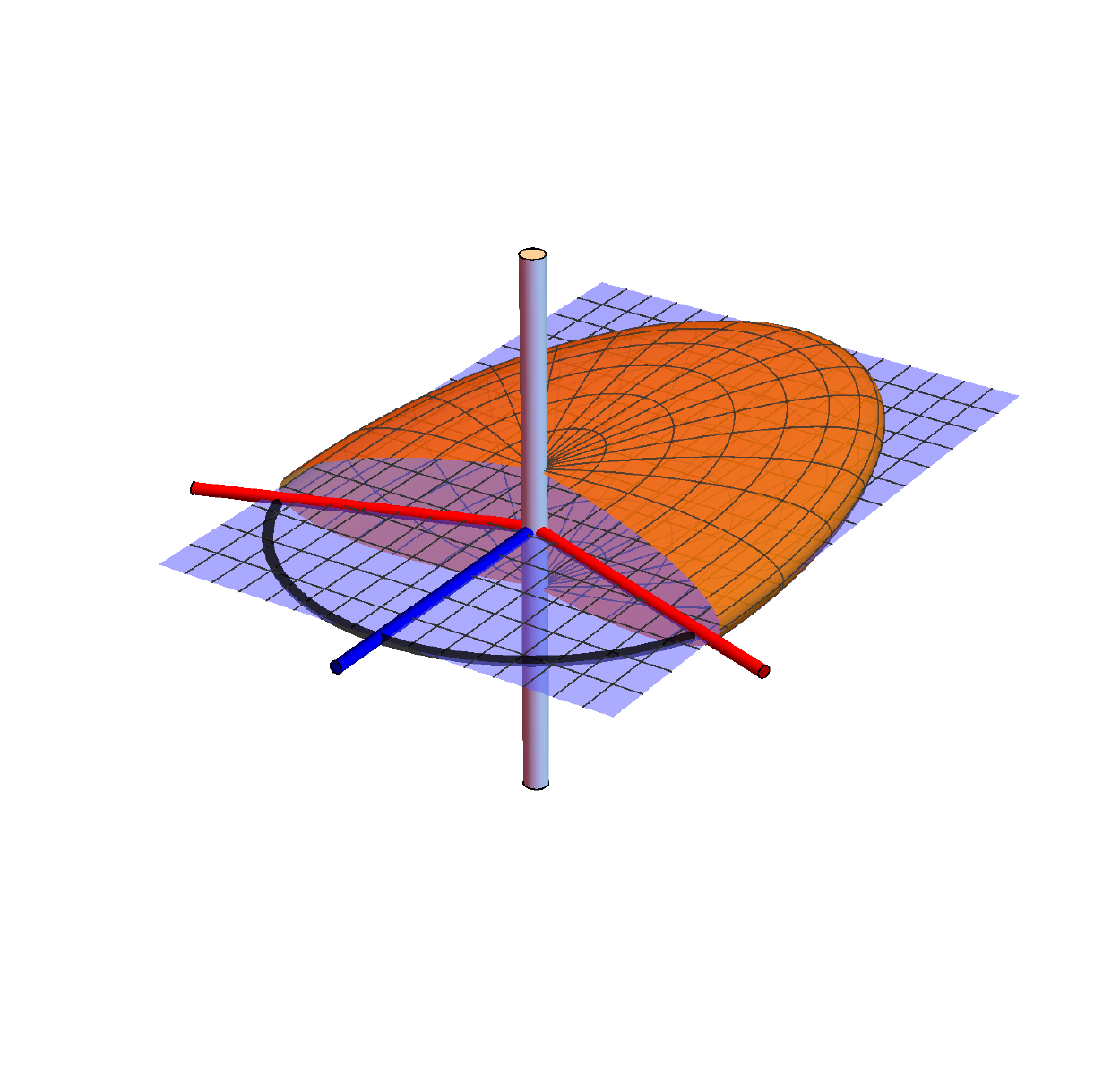}};
    \node[above] at (-2,0.5) {$-\hat{\beta}_{1,0}$};
    \draw[->] (-1.8,0.5) -- (-1, -0.25);
    \node[above] at ( 1.8, 1.3) {$\hat{\beta}_{1,0}$};
    \draw[->] (1.8, 1.3) -- (-0.1, -0.6);
    \begin{scope}[shift={(2,-1.25)},scale=0.5]
      \draw [thick,->] (0,0) -- (0.95 * -0.975, 0.95 * -0.7);
      \node[anchor=north] at (0.95 * -0.975, 0.95 * -0.7) {$q_1$};
      \draw [thick,->] (0,0) -- (1.2  * 1, 1.2 * -0.275);
      \node[anchor=north west] at (1.0  * 1, 1.0 * -0.275) {$q_2$};
      \draw [thick,->] (0,0) -- (0.025, 1.2);
      \node[anchor=south] at (0.025, 1.2) {$q_3$};
    \end{scope}
  \end{tikzpicture}
\end{center}
\captionof{figure}{\label{fig:pms}Action of $\rpolar^\pm_{1,0}(F)$ in
  axes of principal stretch for a stretch ellipsoid with half-axes
  $(\nu_1,\nu_2,\nu_3) = (4,2,1/2)$. The plane of maximal stretch
  $\mathrm{P}^{\rm mp}(F)$ is depicted in blue. The cylinder along
  $q_3 \perp \mathrm{P}^{\rm mp}(F)$ illustrates that the axis
  of rotation is the eigenvector $q_3$ of $U$ associated with the
  smallest singular value $\nu_3 = 1/2$ of $F$. The thin cylinder
  [blue] bisecting the opening represents the relative rotation angle
  $\hat{\beta} = 0$ and corresponds to $\polar(F)$. The outer two
  cylinders [red] correspond to the two non-classical minimizers
  $\rpolar_{1,0}^\pm(F)$. The enclosed angles
  $\hat{\beta}_{1,0}^\pm = \pm\arccos(\frac{2}{\nu_1 + \nu_2})$ are
  the optimal relative rotation angles. This reveals the major
  symmetry of the non-classical minimizers.
}

\begin{rem}[$\rpolar^\pm_{1,0}(F)$ in the classical domain]
  \label{rem:rpolar_class}
  For $\smp(F) \leq 0$ the maximal mean planar strain is non-expansive.
  By definition, we have $F \in \domc_{1,0}$ in the
  classical domain, for which the energy-minimizing relative rotation
  is given by $\widehat{R}_{1,0}(F) = \id$ and there is no deviation
  from the polar factor. In short $\rpolar_{1,0}^\pm(F) = \polar(F)$.
\end{rem}

Let us now turn to the more interesting non-classical case $F \in \domn_{1,0}$.
\begin{rem}[$\rpolar^\pm_{1,0}(F)$ in the non-classical domain]
  \label{rem:rpolar_nonclass}
  If $F \in \domn_{1,0}$, then by definition $\smp(F) > 0$
  and the maximal mean planar strain is expansive. The deviation of
  the non-classical energy-minimizing rotations $\rpolar^\pm_{1,0}(F)$
  from the polar factor $\polar$ is measured by a rotation
  in the plane of maximal stretch $\mathrm{P}^{\rm mp}(F)$ given
  by $\polar(F)^T\rpolar^{\pm}_{1,0}(F) = Q(F)\widehat{R}_{1,0}^\mp(F)Q(F)^T$.
  The rotation axis is the eigenvector $q_3$ associated with the smallest
  singular value $\nu_3 > 0$ of $F$ and the relative rotation angle is
  given by $\hat{\beta}_{1,0}^\mp(F) = \mp\arccos\left(1/\ump(F)\right)$.
  The rotation angles increase monotonically towards the asymptotic limits
  $$
  \lim_{\ump(F) \,\to\, \infty} \hat{\beta}_{1,0}^\pm(F)
  \quad=\quad
  \pm \frac{\pi}{2}\;.
  $$
  In axis-angle representation, we obtain
  \begin{align}
    \widehat{R}_{1,0}^\pm(F) &\quad\equiv\quad \left[\pm \arccos(1/\ump(F)),\, (0,\,0,\,1)\right]\,,\quad\text{and}\\
    \polar^T\rpolar^{\pm}_{1,0}(F) &\quad\equiv\quad \left[\mp \arccos(1/\ump(F)),\, q_3 \right]\;.
  \end{align}
\end{rem}

\begin{cor}[An explicit formula for $\rpolar_{\mu,\mu_c}^\pm(F)$]
  \label{cor:rpolar_formula}
  For the non-classical limit case $(\mu,\mu_c) = (1,0)$ we have
  the following formula for the energy-minimizing Cosserat rotations:
  \begin{equation}
    \rpolar^{\pm}_{1,0}(F)
    \quad\eqdef\quad
    \begin{cases}
      \;\polar(F) &, \text{if}\quad F \in \domc_{1,0}\;,\\
      \;\polar(F)Q(F)\widehat{R}_{1,0}^\mp(F)Q(F)^T &, \text{if}\quad F \in \domn_{1,0}\;.
    \end{cases}
  \end{equation}
  For general values of the weights in the non-classical range
  $\mu > \mu_c \geq 0$, we obtain
  \begin{equation}
    \rpolar^{\pm}_{\mu,\mu_c}(F) \eqdef \rpolar^{\pm}_{1,0}(\widetilde{F}_{\mu,\mu_c})\;,
  \end{equation}
  where $\widetilde{F}_{\mu,\mu_c} \eqdef \lambda^{-1}_{\mu,\mu_c}\,F$ is obtained
  by rescaling the deformation gradient with the inverse of the \emph{induced scaling
  parameter} $\lambda_{\mu,\mu_c} \eqdef \frac{\mu}{\mu - \mu_c} > 0$.
\end{cor}
Note that the previous definition is relative to a fixed choice
of the orthonormal factor $Q(F) \in \SO(3)$ in the spectral
decomposition of $U = QDQ^T$. Further, right from their variational
characterization, one easily deduces that the energy-minimizing
rotations satisfy
$\rpolar^\pm_{\mu, \mu_c}(Q\,F) = Q\,\rpolar^\pm_{\mu,\mu_c}(F)$,
for any $Q \in \SO(3)$, i.e., they are objective
functions;~cf.\remref{rem:polar_vs_rpolar}.

\begin{figure}
  \begin{center}
    \includegraphics[width=12cm]{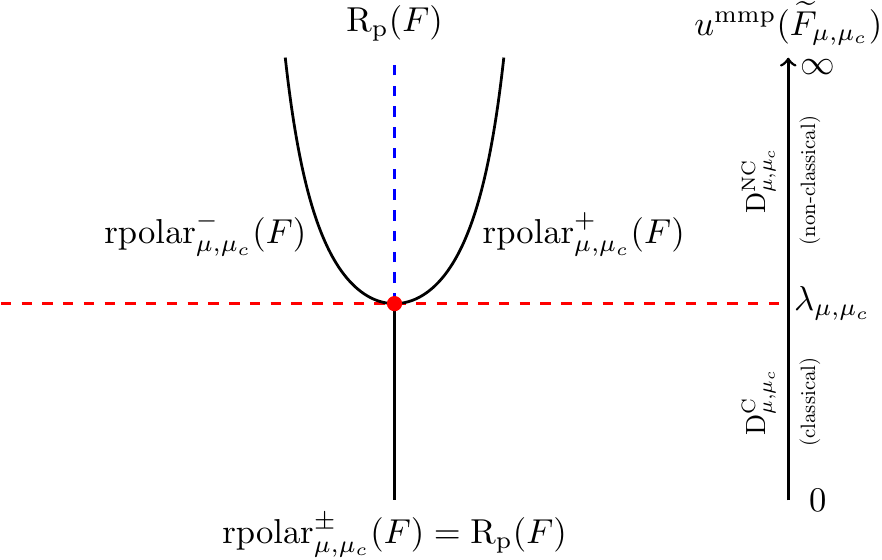}
  \end{center}
  \caption[Pitchfork bifurcation diagram for $\rpolar_{\mu,\mu_c}^\pm(F)$]{
    \label{fig:branchDiag}    Pitchfork bifurcation diagram for $\rpolar_{\mu,\mu_c}^\pm(F)$
    for $\mu > \mu_c \geq 0$. Let us express the energy-minimizers
    $\rpolar_{\mu,\mu_c}^\pm(F)$ in terms of the maximal mean
    planar stretch $\ump(\widetilde{F}_{\mu,\mu_c})$ of the rescaled
    deformation gradient $\widetilde{F}_{\mu,\mu_c} \eqdef \lambda_{\mu,\mu_c}^{-1}F$.
    For values $F \in \domc_{\mu,\mu_c}$, we have
    $0 < \ump \leq \lambda_{\mu,\mu_c}$ and the polar factor
    $\polar(F)$ is uniquely energy-minimizing. In contrast,
    for $F \in \domn_{\mu,\mu_c}$, $\lambda_{\mu,\mu_c} \leq \ump < \infty$, there
    are two non-classical minimizers $\rpolar^\pm_{\mu,\mu_c}(F)$. In this
    regime, the polar factor is no longer optimal but it is still a
    critical point. At the branching point
    $\ump(\widetilde{F}_{\mu,\mu_c}) = \lambda_{\mu,\mu_c}$
    the minimizers all coincide: $\rpolar^{-}_{\mu,\mu_c}(F) = \polar(F) = \rpolar^{+}_{\mu,\mu_c}(F)$. For $\mu_c \to \mu$, the branching point
    escapes to infinity which asymptotically recovers the behavior
    in the classical parameter range $\mu_c \geq \mu > 0$.}
\end{figure}

The domains of the piecewise definition of $\rpolar_{1,0}^\pm(F)$
in~\coref{cor:rpolar_formula} indicate a certain tension-compression
asymmetry in the material model characterized by the Cosserat
shear-stretch energy $W_{1,0}(R\,;F)$. We can also make a second
important observation. To this end, consider
a smooth curve $F(t):(-\epsilon,\epsilon) \to \GL^+(3)$. If the
eigenvector $q_3(t) \in \S^2$ associated with the smallest singular
value $\nu_3(t)$ changes its orientation along this curve, then
the rotation axis of $\rpolar_{1,0}^\pm(F)$ flips as well. Effectively,
the sign of the relative rotation angle $\hat{\beta}_{1,0}^\pm(F)$ is
negated which may lead to jumps.  This can happen, e.g., if $F(t)$
passes through a deformation gradient with a non-simple singular
value, but it may also depend on details of the specific algorithm
used for the computation of the eigenbasis.

For the classical range $\mu_c \geq \mu > 0$, the
polar factor and the relaxed polar factor(s) coincide and trivially
share all properties. This is no longer true for the non-classical parameter
range $\mu_c \geq \mu > 0$ and we compare the properties for that range
in our next remark. More precisely, we present a detailed comparison
of the well-known features of the polar factor $\polar$ which are of
fundamental importance in the context of mechanics.

\begin{minipage}{\linewidth}
  \begin{rem}[$\polar(F)$ vs. $\rpolar(F)$ for the non-classical
      \label{rem:polar_vs_rpolar}
    range $\mu > \mu_c \geq 0$] Let $n \geq 2$ and $F \in \GL^+(n)$.
  The polar factor $\polar(F) \in \SO(n)$ obtained from the polar decomposition
  $F = \polar(F)\,U$ is \emph{always unique} and satisfies:
  \begin{equation}
    \begin{matrix*}[l]
      &\text{(Objectivity)}
      &\hspace{.125\linewidth} &\polar(\,Q\cdot F\,)  &= &Q\cdot\polar(F)   &\quad\quad(\forall Q \in \SO(n))\;,\\
      &\text{(Isotropy)}
      &\hspace{.125\linewidth} &\polar(\,F\cdot Q\,)  &= &\polar(F)\cdot Q   &\quad\quad(\forall Q \in \SO(n))\;,\\
      &\text{(Scaling invariance)}
      &\hspace{.125\linewidth} &\polar(\,\lambda\cdot F\,) &= &\polar(F)  &\quad\quad(\forall \lambda > 0)\;,\\
      &\text{(Inversion symmetry)}
      &\hspace{.125\linewidth}                     &\polar(F^{-1}) &= &\polar(F)^{-1}\;. &
    \end{matrix*}
  \end{equation}
  The relaxed polar factor(s) $\rpolar_{\mu,\mu_c}(F) \subset \SO(n)$ is
  \emph{in general multi-valued} and, due to its variational
  characterization, satisfies:
  \begin{equation}
    \begin{matrix*}[l]
      &\text{(Objectivity)}
      &\hspace{.125\linewidth} & \rpolar_{\mu,\mu_c}(\,Q \cdot F\,)    &= &Q \cdot \rpolar_{\mu,\mu_c}(F)    &\quad\quad(\forall Q \in \SO(n))\;,\\
      &\text{(Isotropy)}
      &\hspace{.125\linewidth} &\rpolar_{\mu,\mu_c}(\,F\cdot Q\,)     &= &\rpolar_{\mu,\mu_c}(F)\cdot Q    &\quad\quad(\forall Q \in \SO(n))\;.
    \end{matrix*}
  \end{equation}
  For the particular dimensions $k = 2,3$, our explicit formulae imply
  that there exist particular instances $\lambda^* > 0$
  and $F^* \in \GL^+(k)$, for which we have
  \begin{equation}
    \begin{matrix*}[l]
      & \text{(\underline{Broken} scaling invariance)}
      &\hspace{.125\linewidth}
      &\rpolar^\pm_{\mu,\mu_c}(\lambda^* \cdot F^*) &\neq &\rpolar(F^*)
      &, \quad\text{and}\\
      &\text{(\underline{Broken} inversion symmetry)}
      &\hspace{.125\linewidth}
      &\rpolar^\pm_{\mu,\mu_c}({F^*}^{-1})  &\neq &\rpolar(F^*)^{-1} &.
    \end{matrix*}
  \end{equation}
  This can be directly inferred from the partitioning of
  $\GL^+(k) = \domc_{\mu,\mu_c} \,\cup\, \domn_{\mu,\mu_c}$
  and the respective piecewise definition of the relaxed
  polar factor(s), see~\coref{cor:rpolar_formula}.
\end{rem}
We interpret these broken symmetries as a (generalized) tension-compression
asymmetry.
\end{minipage}

\subsection{The reduced Cosserat shear-stretch energy}
We now introduce the notion of a reduced energy as the energy level
realized by the energy-minimizing rotations $\rpolar_{\mu,\mu_c}(F)$.
\begin{defi}[Reduced Cosserat shear-stretch energy]
  The \textbf{reduced Cosserat shear-stretch energy} is defined as
  \begin{equation}
    W_{\mu,\mu_c}^{\rm red}: \GL^+(n) \to \RPosZ,
    \quad\quad
    W_{\mu, \mu_c}^{\rm red}(F) \;\eqdef\; \min_{R\,\in\,\SO(n)} W_{\mu,\mu_c}(R\,;F)\;.
  \end{equation}
\end{defi}
Besides the previous definition, we also have the following equivalent
means for the explicit computation of the reduced energy
\begin{equation}
  \begin{aligned}
    W_{\mu,\mu_c}^{\rm red}(F) &\;=\; W_{\mu,\mu_c}(\rpolar^\pm_{\mu,\mu_c}(F)\,;F)\;,\quad\text{and}\\
    W_{\mu,\mu_c}^{\rm red}(F)
    &\;=\;
    W_{\mu, \mu_c}^{\rm red}(D)
    \;\eqdef
    \min_{\widehat{R} \in \SO(n)} W_{\mu,\mu_c}(\widehat{R}\,; D)
    \;=\;
    W_{\mu,\mu_c}(\widehat{R}_{\mu,\mu_c}^\pm\,;D)\;.
  \end{aligned}
\end{equation}

\begin{lem}[The reduced Cosserat shear-stretch energy $\wsymred(F)$ in terms of singular values]
  \label{lem:wred10_singular}
  Let $F \in \GL^+(3)$ and $\nu_1 > \nu_2 > \nu_3 > 0$ the
  ordered singular values of $F$. Then the reduced Cosserat shear-stretch
  energy $\wsymred(F)$ admits the following piecewise representation
  \begin{equation*}
    \wsymred(F) \,=\, \begin{cases}
      \,(\nu_1 - 1)^2 + (\nu_2 - 1)^2 + (\nu_3 - 1)^2 = \hsnorm{U - \id}^2
      &,\,\text{if}\;\,\nu_1 + \nu_2 \leq 2,\,\text{i.e.},\, F \in \domc_{1,0}\,,\\
      \,\frac{1}{2}\,(\nu_1 - \nu_2)^2 + (\nu_3 - 1)^2
      &,\,\text{if}\;\, \nu_1 + \nu_2 \geq 2,\,\text{i.e.},\,F \in \domn_{1,0}\,.\\
    \end{cases}
  \end{equation*}
\end{lem}
Our next step is to reveal the form of the reduced energy for the entire
non-classical parameter range $\mu > \mu_c \geq 0$ which involves the
parameter reduction lemma, but we have to be a bit careful.

\begin{rem}[Reduced energies and the parameter reduction lemma]
  The parameter reduction in~\leref{lem:parameter_reduction} is
  the key step in the computation of the minimizers for general
  non-classical material parameters $\mu > \mu_c \geq 0$.
  It might be tempting, but we have to stress that the general form of
  the reduced energy cannot be obtained by rescaling the singular values
  $\nu_i \mapsto \lambda_{\mu,\mu_c}^{-1}\nu_i$ in the singular value
  representation of $W^{\rm red}_{1,0}$.
\end{rem}

\begin{theo}[$W^{\rm red}_{\mu,\mu_c}$ as a function of the singular values]
\label{theo:wmm_explicit}
Let $F \in \GL^+(n)$ and $\nu_1 > \nu_2 > \nu_3 > 0$, the ordered
singular values of $F$ and let $\mu > \mu_c \geq 0$, i.e., a non-classical
parameter set. Then the reduced Cosserat shear-stretch energy
$W^{\rm red}_{\mu,\mu_c}: \GL^+(3) \to \RPosZ$
admits the following explicit representation
\begin{align*}
  W^{\rm red}_{\mu,\mu_c}(F) \,=\,
  \begin{cases}
    \, \mu \left((\nu_1 - 1)^2 + (\nu_2 - 1)^2 + (\nu_3 -
    1)^2\right) = \mu\,\hsnorm{U - \id}^2
    &,\; F \in \domc_{\mu,\mu_c}\;,\\
    \, \frac{\mu}{2}(\nu_1 - \nu_2)^2
    + \mu\, (\nu_3 - 1)^2
    + \frac{\mu_c}{2}\left(\left(\nu_1 + \nu_2\right) - \sradmm\right)^2
    - \frac{\mu_c}{2}\cdot\rho_{\mu,\mu_c}^2
    &,\; F \in \domn_{\mu,\mu_c}\;.
  \end{cases}
\end{align*}
\end{theo}
\begin{rem}[On $\mu_c$ as a penalty weight]
Let us consider the contribution of the skew-term to
$W^{\rm red}_{\mu,\mu_c}$
given by
$$
\frac{\mu_c}{2}\left(\left(\nu_1 + \nu_2\right) - \sradmm\right)^2
$$
as a penalty term for $F \in \GL^+(3)$ arising for material parameters
in the non-classical parameter range $\mu > \mu_c \geq 0$. This leads to
a simple but interesting observation for strictly positive $\mu_c > 0$.
The minimizers $F \in \GL^+(3)$ \emph{for the penalty term} satisfy
the bifurcation criterion
$$\nu_1 + \nu_2 = \sradmm$$
for $\rpolar^\pm_{\mu,\mu_c}(F)$. In this case
$\widehat{R}_{\mu,\mu_c}^\pm = \id$ which implies that
$\widehat{R}_{\mu,\mu_c}^\pm D - \id \in \Sym(3)$, i.e.,
it is symmetric. Hence, the skew-part vanishes entirely which
minimizes the penalty. In numerical applications, a rotation
field $R$ approximating $\rpolar^\pm(F)$ can be expected to
be unstable in the vicinity of the branching point
$\nu_1 + \nu_2 \approx \sradmm$.
Hence, a penalty which explicitly rewards an approximation to the
bifurcation point seems to be a delicate property. In strong contrast,
for the case when the Cosserat couple modulus is zero, i.e.,
$\mu_c = 0$, the penalty term vanishes entirely. This hints at a
possibly more favorable qualitative behavior of the model in that
case; cf.~\cite{Neff_ZAMM05}.
\end{rem}

We recall that the tangent bundle $T\SO(n)$ is isomorphic to the product
$\SO(n)\times\so(n)$ as a vector bundle. This is commonly referred
to as the left trivialization, see, e.g.,~\cite{Duistermaat:2012:LG}.
With this we can minimize over the tangent bundle in the following
\begin{lem}
\label{lem:dist_ctb_technical}
Let $F \in \R^{n\times n}$. Then
\begin{align*}
  \inf_{\substack{R\,\in\,\SO(n)\\ A\,\in\,\so(n)}} \norm{R^TF - \id - A}^2 \quad=\quad \min_{R\,\in\,\SO(n)} \norm{\sym(R^TF - \id)}^2 \quad \defeq \quad \min_{R\,\in\,\SO(n)} W_{1,0}(R\,;F) \;.
\end{align*}
\end{lem}
In the non-classical limit case $(\mu,\mu_c) = (1,0)$, the preceding
lemma yields a geometric characterization of the \emph{reduced}
Cosserat shear-stretch energy as a distance which we find remarkable.
\begin{cor}[Characterization of $W_{1,0}^{\rm red}$ as a distance]
  \label{cor:wred10_distance}
  Let $n \geq 2$ and consider $F \in \GL^+(n)$ with singular values
  $\nu_1 \geq \nu_2 \geq \ldots \geq \nu_n > 0$, i.e., not
  necessarily distinct. Then the reduced Cosserat shear-stretch energy
  $W_{1,0}^{\rm red}: \GL^+(n) \to \RPosZ$
  admits the following characterization as a distance
  \begin{equation}
    W_{1,0}^{\rm red}(F) \quad=\quad \dist_{\rm euclid}^2\big(F,\, \SO(n)\left(\id + \so(n)\right)\big)\;.
  \end{equation}
  Here, $\dist_{\rm euclid}$ denotes the euclidean distance function.
\end{cor}

\begin{figure}[t]
  \begin{center}
    \includegraphics[width=7.2cm,clip=false,trim=3.2cm 2cm 3.2cm 3.2cm]{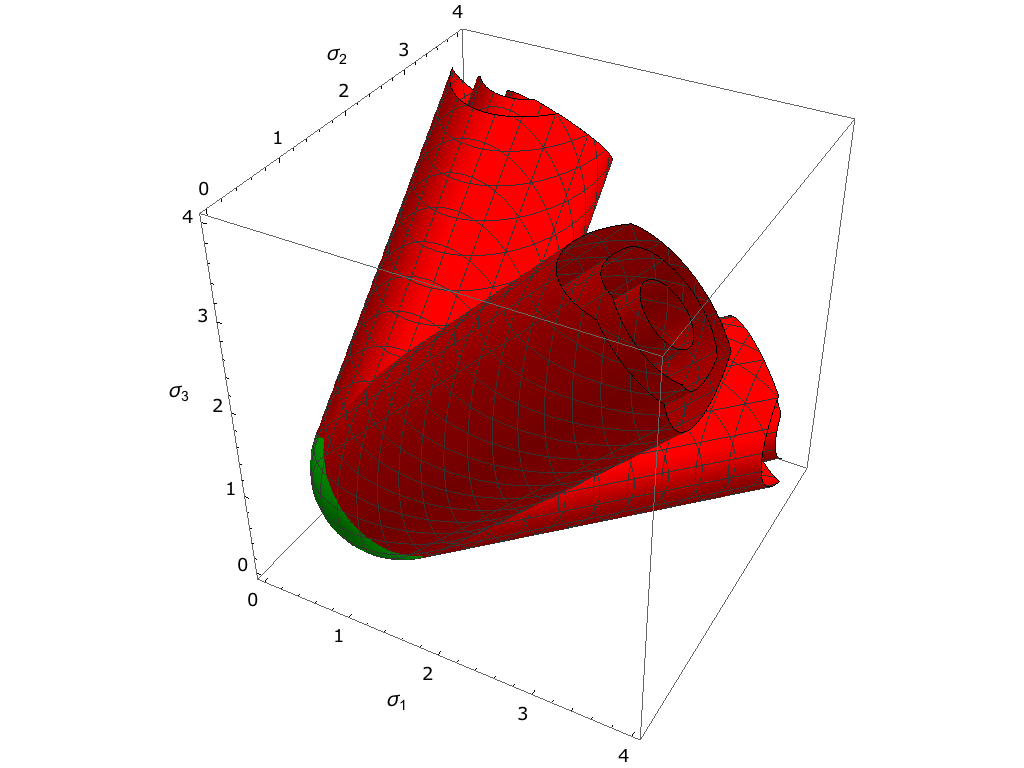}
    \includegraphics[width=7.2cm,clip=false,trim=3.2cm 2cm 3.2cm 3.2cm]{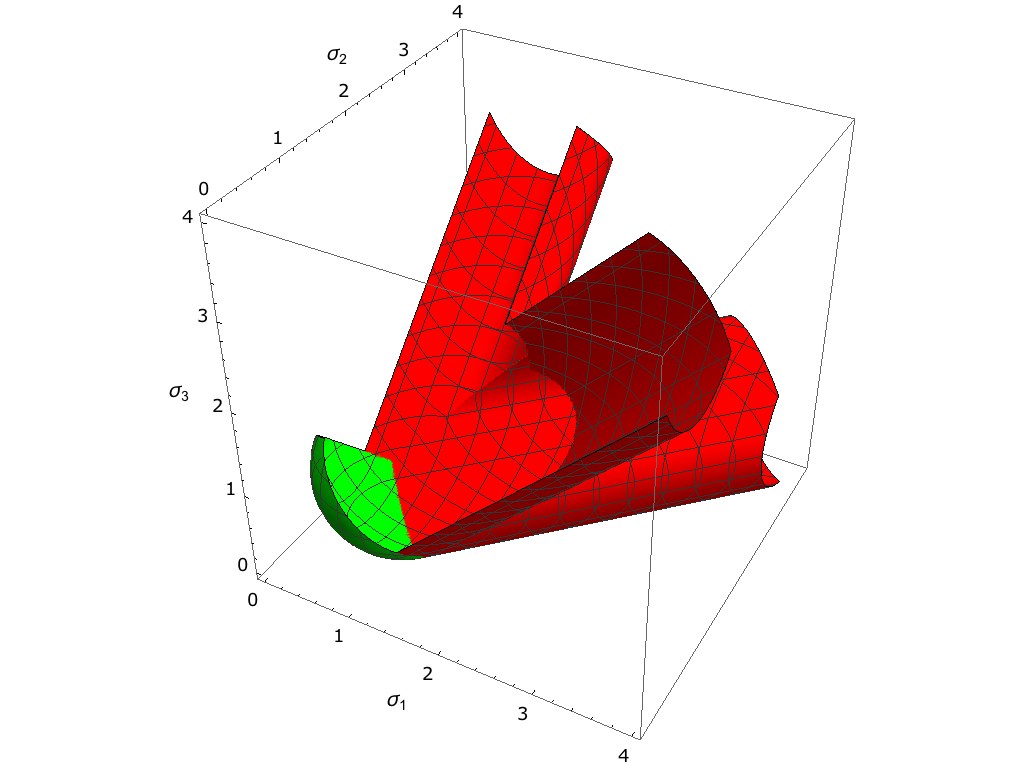}
  \end{center}
  \caption[Energy isosurfaces of {$W^{\rm red}_{1,0}$} in the space of
    singular values of $F$]{\label{fig:wred_isosurf}    Energy isosurfaces of $W^{\rm red}_{1,0}$ considered as
    a function of the \emph{unordered} singular values
    $\nu_1,\nu_2,\nu_3 > 0$ of $F \in \GL^+(3)$.
    The displayed contour levels are $0.1$, $0.4$ and $0.8$.
    On the right, we have removed a piece from the non-classical
    cylindrical parts (red) of the energy level $0.8$ which reveals
    the spherical shell of the classical part (green). Note that
    a computation of these level surfaces via Monte Carlo minimization
    yields the same result (but at a much lower resolution).}
\end{figure}

\subsection{Alternative criteria for the existence of non-classical solutions}
For $\mu > \mu_c > 0$, i.e., for strictly positive $\mu_c > 0$,
the singular radius satisfies $\sradmm \eqdef \frac{2\mu}{\mu - \mu_c} > 2$.
We now define a quite similar constant, namely
\begin{align}
\zeta_{\mu,\mu_c} \eqdef \sradmm - \;\rho_{1,0}
= \frac{2\mu_c}{\mu - \mu_c} > 0\;.
\end{align}
Furthermore, we define the $\epsilon$-neighborhood of a set
$\mathcal{X} \subseteq \R^{n\times n}$ relative to the euclidean
distance function as
$$
N_{\epsilon}(\mathcal{X})
\;\eqdef\;
\setdef{Y \in \R^{n \times n}}{\dist_{\rm euclid}(Y, \mathcal{X}) < \epsilon}\;.
$$

\begin{figure}[h!]
  \begin{center}
    \scalebox{0.75}{
    \begin{tikzpicture}
      \node (Pic) at (0,0)
            {\includegraphics[width=8cm]{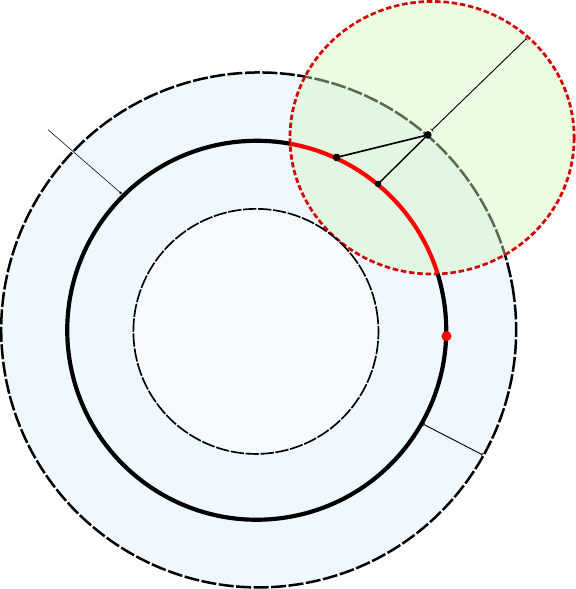}};
            \node[scale=1.2] (Id) at (2.5,-0.55) {$\id$};
            \node[rotate=-33] (UEpsSO3) at (-2.1, -3.05) {$N_\epsilon(\SO(3))$};
            \node[rotate=-25] (HalfEps) at (2.35, -1.85) {$\epsilon$};
            \node[rotate=45] (Eps) at (2.6, 3.1) {$\delta$};
            \node (SO3) at (-3.4,2.6) {$\SO(3)$};
            \node (F) at (2.0, 2.6) {$F$};
            \node (UEpsF) at (3.25,2) {$N_{\delta}(F)$};
            \node (R) at (0.5,1.7) {$R$};
            \node[rotate=12] (HR) at (1.2,2.2) {$F-R$};
    \end{tikzpicture}}
  \end{center}
  \caption{Illustration of a euclidean $\epsilon$-neighborhood of $\SO(3) \subset
    \R^{3\times 3}$.}
\end{figure}

\begin{lem}[Classical $\SO(3)$-neighborhood for $\mu_c > 0$]
  Let $\mu > \mu_c > 0$, $F \in \GL^+(3)$ and
  $\zeta_{\mu,\mu_c} \eqdef \frac{2\mu_c}{\mu - \mu_c} > 0$.
  Then we have the following inclusion
\begin{align}
  N_{\frac{1}{2}\zeta_{\mu,\mu_c}^2}(\SO(3)) \quad\subset\quad \domc_{\mu,\mu_c}\;.
\end{align}
In other words, for all $F \in \GL^+(3)$ satisfying
$\dist_{\rm euclid}(F, \SO(3)) = \hsnorm{U - \id}^2 < \frac{1}{2}\zeta_{\mu,\mu_c}^2$,
the polar factor $\polar{}$ is the unique minimizer of
$W_{\mu,\mu_c}(R\,;F)$.
\end{lem}

\begin{lem}
\label{lem:disc:SL3_dom_nc}
Let $F \in \SL(3)$, i.e., $\det{F} = \nu_1\nu_2\nu_3 = 1$,
where $\nu_1 \geq \nu_2 \geq \nu_3 > 0$ are ordered
singular values of $F$, not necessarily distinct. Then
\begin{align}
  \SL(3) \quad\subset\quad \domn_{1,0}\;,
\end{align}
i.e., $F$ induces a strictly non-classical minimizer. Equivalently,
$\det{F} = 1$ implies the estimate $\nu_1 + \nu_2 \geq 2$.
\end{lem}

\begin{rem}
  \label{rem:disc:SL3_dom_nc_strict}
  If we make the stronger assumption $\nu_1 > \nu_2 > \nu_3 > 0$,
  we obtain a strict inequality $\nu_1 + \nu_2 > 2$. In that case,
  $F \in \domn_{1,0} \setminus \domc_{1,0}$ is strictly non-classical.
\end{rem}

\begin{cor}
  \label{cor:SL3_dom_nc_strict}
  Let $\mu > 0$,
    $F \in \SL^+(3)$ and assume
  that $\nu_1 > \nu_2 > \nu_3 > 0$. Then
  \begin{equation}
    F \quad\in\quad \domn_{\mu,0}\setminus \domc_{\mu,0}\;,
  \end{equation}
  i.e., the minimizers $\rpolar_{\mu,0}^\pm(F) \neq \polar{}$ are \emph{strictly}
  non-classical.
\end{cor}

\begin{figure}
  \parbox{0.8\textwidth}{
    \begin{tikzpicture}
      \node[anchor=west] (Pic) at (0,0)
           {\includegraphics[width=11.0cm]{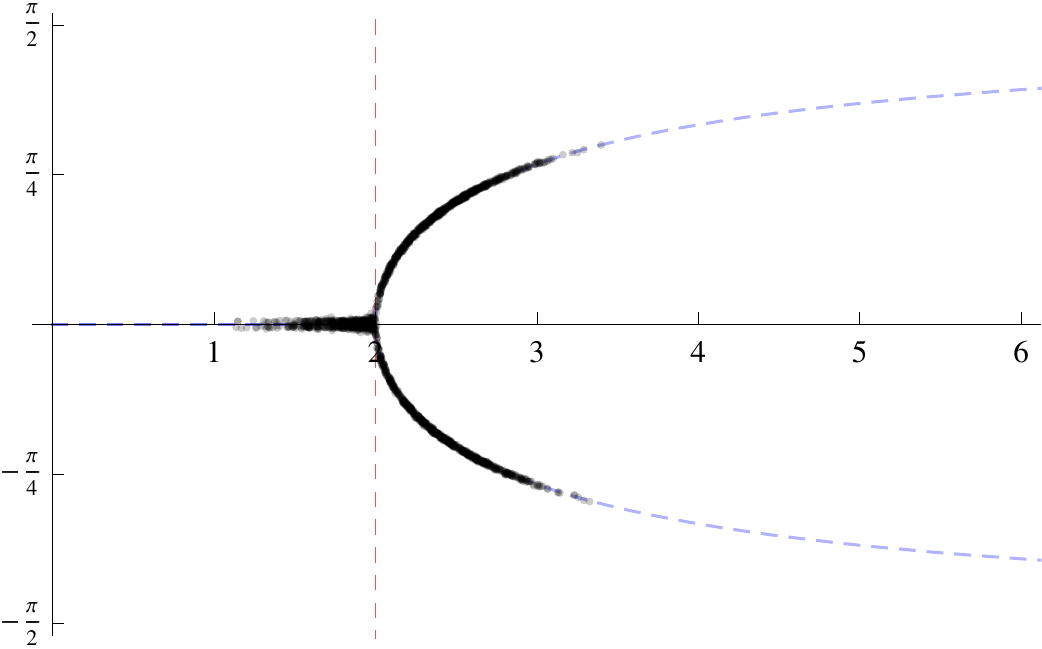}};
            \fill [white] (0.1, 3.5) rectangle (0.6, 1.0);
      \fill [white] (0.1,-1.0) rectangle (0.6,-3.5);
            \node [anchor=east] at (0.7,  pi/2) {$\frac{\pi}{4}$};
      \node [anchor=east] at (0.7,  pi)   {$\frac{\pi}{2}$};
      \node [anchor=east] at (0.7, -pi/2) {$-\frac{\pi}{4}$};
      \node [anchor=east] at (0.7, -pi)   {$-\frac{\pi}{2}$};
            \foreach \x in {1,3,4,5,6}
      {
        \fill [white] (.675 + 1.705*\x - 0.15, -0.1) rectangle (.675 + 1.705*\x + 0.15, -0.5);
        \node at (.675 + 1.705*\x, -0.3) {$\x$};
      }
      \node [anchor=east] (rho) at (5.6, -2.5) {$\rho_{1,0} = 2$};
  \end{tikzpicture}}%
  \parbox{0.2\textwidth}{%
    Parameters: $(\mu,\mu_c) = (1,0)$\\
    $\rho_{1,0} = 2$
  }
  \parbox{0.8\textwidth}{
    \begin{tikzpicture}
      \node[anchor=west] (Pic) at (0,0)
           {\includegraphics[width=11.0cm]{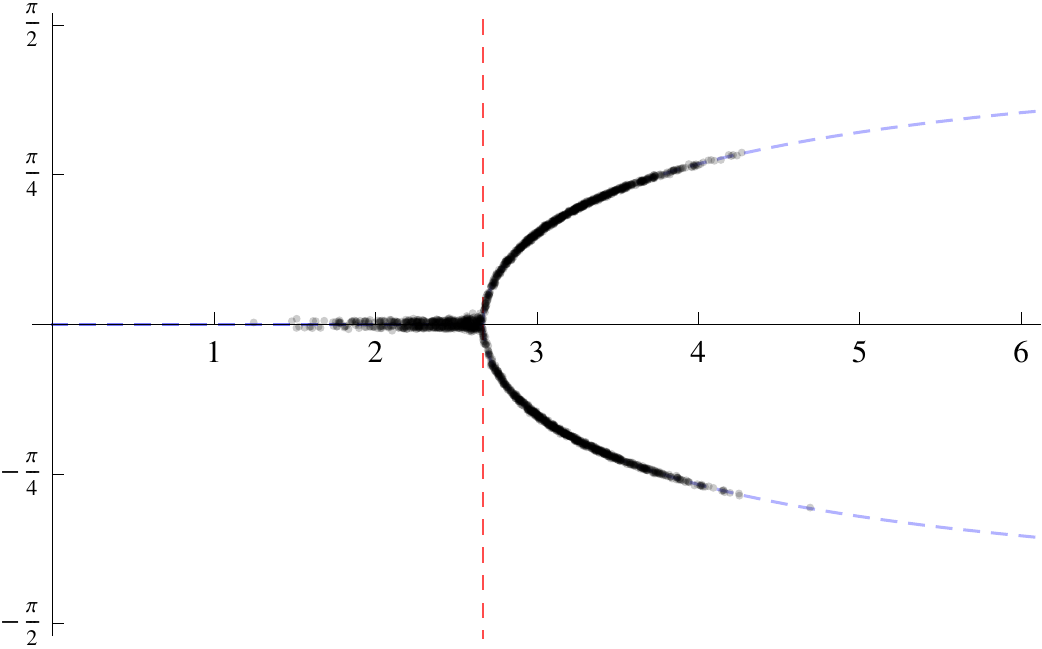}};
            \fill [white] (0.1, 3.5) rectangle (0.6, 1.0);
      \fill [white] (0.1,-1.0) rectangle (0.6,-3.5);
            \node [anchor=east] at (0.7,  pi/2) {$\frac{\pi}{4}$};
      \node [anchor=east] at (0.7,  pi)   {$\frac{\pi}{2}$};
      \node [anchor=east] at (0.7, -pi/2) {$-\frac{\pi}{4}$};
      \node [anchor=east] at (0.7, -pi)   {$-\frac{\pi}{2}$};
            \foreach \x in {1,2,3,4,5,6}
      {
        \fill [white] (.675 + 1.705*\x - 0.15, -0.1) rectangle (.675 + 1.705*\x + 0.15, -0.5);
        \node at (.675 + 1.705*\x, -0.3) {$\x$};
      }
      \node (rho) at (6.1, -2.5) {$\rho_{1,\frac{1}{4}} = \frac{8}{3}$};
  \end{tikzpicture}}%
  \parbox{0.2\textwidth}{%
    Parameters: $(\mu,\mu_c) = (1,\frac{1}{4})$\\
    $\rho_{1,\frac{1}{4}} = \frac{8}{3}$
  }
  \parbox{0.8\textwidth}{
    \begin{tikzpicture}
      \node[anchor=west] (Pic) at (0,0)
            {\includegraphics[width=11.0cm]{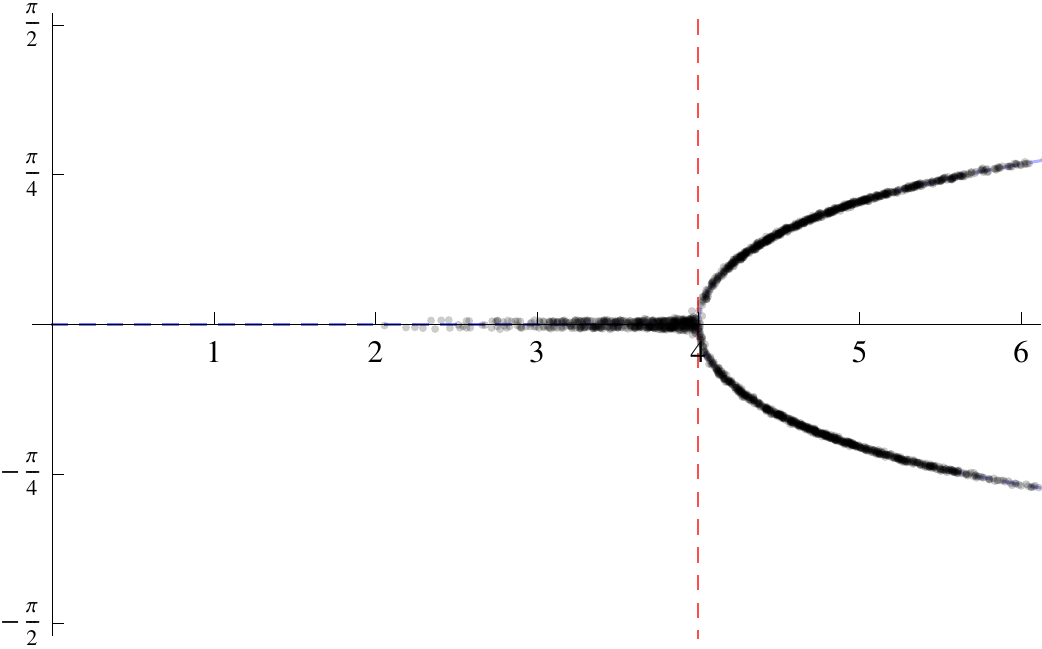}};
            \fill [white] (0.1, 3.5) rectangle (0.6, 1.0);
      \fill [white] (0.1,-1.0) rectangle (0.6,-3.5);
            \node [anchor=east] at (0.7,  pi/2) {$\frac{\pi}{4}$};
      \node [anchor=east] at (0.7,  pi)   {$\frac{\pi}{2}$};
      \node [anchor=east] at (0.7, -pi/2) {$-\frac{\pi}{4}$};
      \node [anchor=east] at (0.7, -pi)   {$-\frac{\pi}{2}$};
            \foreach \x in {1,2,3,5,6}
      {
        \fill [white] (.675 + 1.705*\x - 0.15, -0.1) rectangle (.675 + 1.705*\x + 0.15, -0.5);
        \node at (.675 + 1.705*\x, -0.3) {$\x$};
      }
      \node (rho) at (8.35, -2.5) {$\rho_{1,\frac{1}{2}} = 4$};
  \end{tikzpicture}}%
  \parbox{0.2\textwidth}{%
    Parameters: $(\mu,\mu_c) = (1, \frac{1}{2})$\\
    $\rho_{1,\frac{1}{2}} = 4$
  }
  \caption[Optimal non-classical relative rotation angles]{\label{fig:hat{beta}MC_NC}
    Optimal relative rotation angles $\hat{\beta}^{\rm MC}_{\mu,\mu_c}$ for
    multiple non-classical values $\mu > \mu_c \geq 0$. The angles are
    obtained by stochastic (Monte Carlo) minimization of $W_{\mu,\mu_c}(R\,;F)$.
    The dashed blue curve shows the predicted value for
    $\hat{\beta}_{1,0}^\pm(\nu_1 + \nu_2)$ and
    the dashed red line marks the expected bifurcation point
    at $\rho_{\mu,\mu_c}$. For a direct comparison, we
    provide~\figref{fig:hat{beta}MC_C} on
    page~\pageref{fig:hat{beta}MC_C} which shows the classical
    limit case $(\mu,\mu_c) = (1,1)$; see also~\figref{fig:branchDiag}
    on page~\pageref{fig:branchDiag} for an illustration and a more
    precise description of the bifurcation behavior predicted by
    our proposed formula~$\rpolar^\pm_{\mu,\mu_c}(F)$.}
\end{figure}
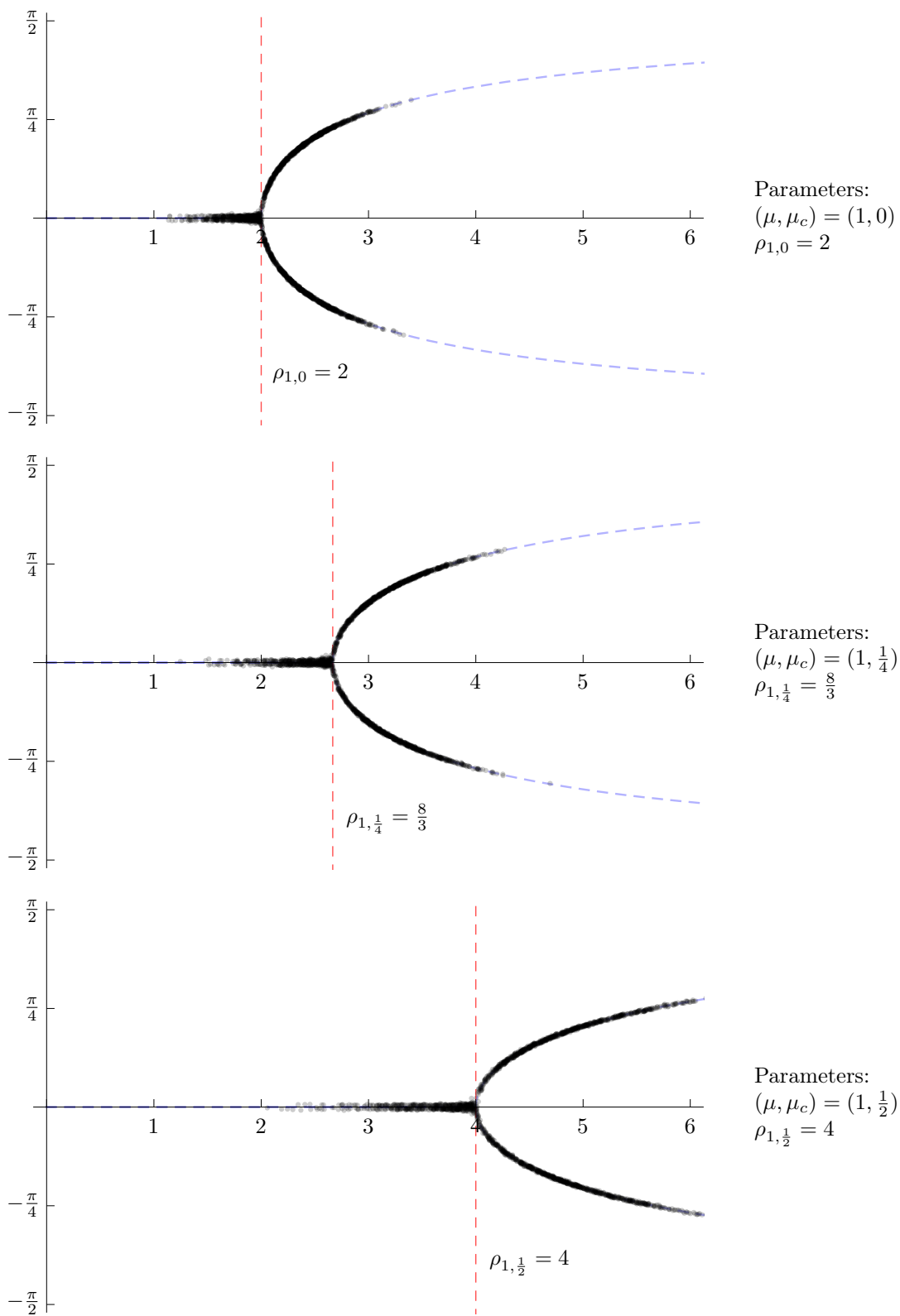

\begin{figure}
  \parbox{0.8\textwidth}{
    \begin{tikzpicture}
      \node[anchor=west] (Pic) at (0,0)
           {\includegraphics[width=11cm]{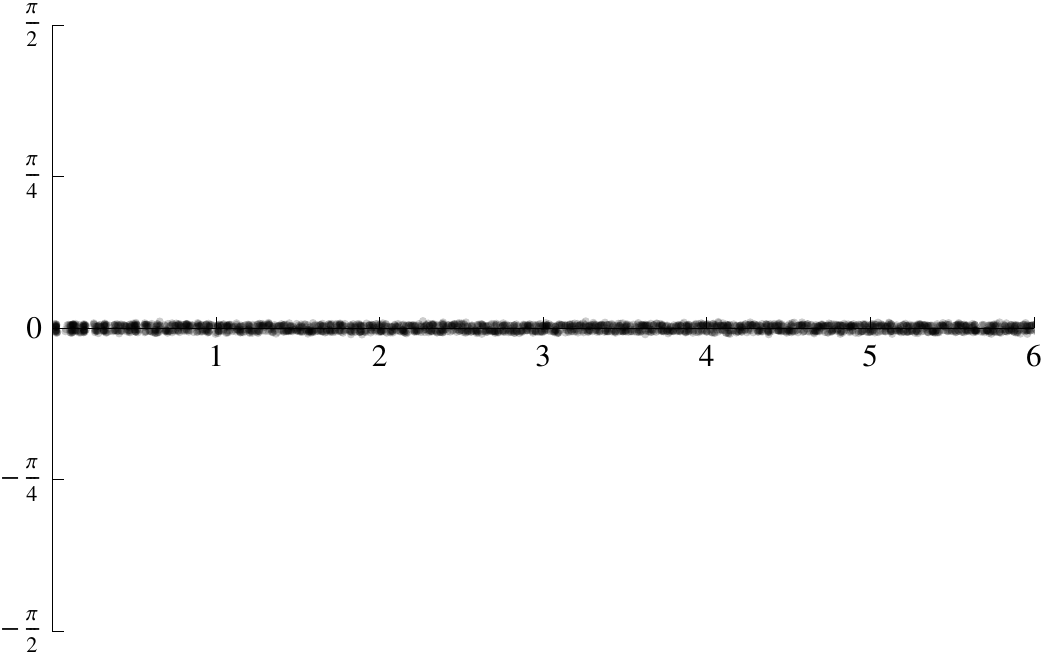}};
      %% Overdraw broken y-tick labels with tikz
      \fill [white] (0.1, 3.5) rectangle (0.6, 1.0);
      \fill [white] (0.1,-1.0) rectangle (0.6,-3.5);
      %% Print new y-tick labels
      \node [anchor=east] at (0.7, pi/2) {$\frac{\pi}{4}$};
      \node [anchor=east] at (0.7, pi)   {$\frac{\pi}{2}$};
      \node [anchor=east] at (0.7,-pi/2) {$-\frac{\pi}{4}$};
      \node [anchor=east] at (0.7,-pi)   {$-\frac{\pi}{2}$};
      %% Overdraw broken x-tick labels
      \foreach \x in {1,2,3,5,6}
      {
        \fill [white] (.675 + 1.725*\x - 0.15, -0.1) rectangle (.675 + 1.725*\x + 0.15, -0.5);
        \node at (.675 + 1.725*\x, -0.3) {$\x$};
      }
  \end{tikzpicture}}%
  \parbox{0.2\textwidth}{%
    Parameters: $(\mu,\mu_c) = (1, 1)$\\
    $\rho_{1,1} = \infty$
  }
  \caption[Optimal classical relative rotation angles]{\label{fig:hat{beta}MC_C}%
    Optimal relative rotation angle $\hat{\beta}_{1,1}^{\rm MC}$ obtained
    from stochastic (Monte Carlo) minimization for the classical limit case
    $\mu = \mu_c = 1$. We observe that the relative rotation angle vanishes
    up to numerical accuracy, since the polar factor $\polar(F)$ is always
    optimal in perfect accordance with Grioli's theorem,
    see~\cite{Neff_Grioli14} and~\cite[Cor.~2.4,~p.~5]{Fischle:2016:OC2D}.
    More precisely, this corresponds to the prediction $\hat{\beta}^\pm_{1,1}(\nu_1 + \nu_2) = 0$.}
\end{figure}

 \countres
\section{Optimal rotations in general dimension}
The key insight for the solution of the minimization problem
in general dimension $n \geq 2$ is a new approach to the
analysis of the critical points. The Euler-Lagrange equations for
$W_{1,0}(R\,;F)$ are equivalent to
\begin{equation}
  (\widehat{R}D - \id)^2 \;\in\; \Sym(n)\;.
\end{equation}
This is a \emph{symmetric square condition} for the relative rotation
$\widehat{R}$, since
\begin{equation}
  \left(X(\widehat{R})\right)^2 \;=\; S \in \Sym(n),
  \quad\quad\text{where}\quad\quad
  X(\widehat{R}) \eqdef \widehat{R}D - \id \in \R^{n \times n}\;.
\end{equation}
As it is sufficient to compute the optimal relative rotation
$\widehat{R}$, we simply set $R = \widehat{R}$ for the rest
of this section.\\

One might suspect that the critical points of $W_{1,0}(R\,;D)$
are connected to real matrix square roots of real symmetric
matrices. And indeed, the structure of the set of critical
points of $W_{1,0}(R\,;D)$ can be revealed quite elegantly by a specific
characterization of the set of real matrix square roots of real
symmetric matrices. Note that this characterization~\cite[Thm. 2.13]{Borisov:2016:ORP}, which is similar in spirit to the standard representation
theorem for orthogonal matrices $\O(n)$ as block matrices, seems not
to be known in the literature. Due to this representation, the square
roots of interest can always be orthogonally transformed into a
block-diagonal representation which reduces the minimization problem
from arbitrary dimension $n > 2$ into decoupled one- and two-dimensional
subproblems. These can then be solved independently.
From this point of view, a non-classical minimizer in $n\!=\!3$,
simultaneously solves a one-dimensional and a two-dimensional
subproblem. The one-dimensional problem determines the rotation axis
of the optimal rotations, while the two-dimensional subproblem
determines the optimal rotation angles.

The degenerate cases of optimal Cosserat rotations arising for
recurring parameter values $\nu_i$, $i = 1,2,3$, in the diagonal
parameter matrix $D \in \Diag(n)$ has not been treated previously
in~\cite{Fischle:2016:OC3D}, but is also accessible with the
general approach. Note that this case corresponds to the special
case of two or more equal principal stretches $\nu_i$
which is an important highly symmetric corner case in mechanics.

Combining the results of the two preceding sections, we can now
describe the critical values of the Cosserat shear-stretch energy
$W_{1,0}(R\,;D)$ which are attained at the critical points. The main result
of this section is a procedure (algorithm) which traverses the set of
critical points in a way that reduces the energy at every step of the
procedure and finally terminates in the subset of global minimizers.

Technically, we label the critical points by certain partitions
of the index set $\{1,\ldots,n\}$ containing only subsets $I$ with one
or two elements. In the last section, we have seen that the subsets
$I$ and a choice of sign for $\det{R_I}$ uniquely characterize a critical
point $R \in \SO(n)$.

Let us give an outline of the energy-decreasing traversal strategy
starting from a given labeling partition (i.e., critical point):
\begin{enumerate}
\item Choose the positive sign $\det{R_I} = +1$ for each subset
      of the partition.
\item Disentangle all overlapping blocks for $n > 3$
     (cf.~\leref{lem:overlap}).
\item Successively shift all $2\times2$-blocks to the lowest
      possible index, i.e., collect the blocks of size two
      as close to the upper left corner of the matrix $R$
      as possible (cf.~\leref{lem:comparison}).
\item Introduce as many additional $2\times 2$-blocks by joining
      adjacent blocks of size $1$ as the constraint $\nu_i + \nu_j > 2$
      allows (cf.~\leref{lem:comparison}).
\end{enumerate}

The next theorem connects the value of $W_{1,0}(R\,;D)$ realized by
a critical point with its labeling partition and the choice of
determinants $\det{R_I}$ which characterize it.
\begin{theo}[Characterization of critical points and values]
  \label{theo:critical_values}
  Let the entries $\nu_1>\nu_2>\ldots>\nu_n>0$ of $D \in \Diag(n)$.
  Then the critical points $R \in \SO(n)$ can be classified according
  to partitions of the index set $\{1,\ldots,n\}$ into subsets of size
  one or two and choices of signs for the determinant $\det{R_I}$ for
  each subset $I$. The subsets of size two $I = \{i,j\}$ satisfy
  $$
  \begin{cases}
  \phantom{|} \nu_i + \nu_j \phantom{|} > 2,     & \quad \det{R_I} = +1\;,\quad\text{and}\\
  \abs{\nu_i - \nu_j} > 2, & \quad \det{R_I} = -1\;.
  \end{cases}
  $$
  The corresponding critical values are given by
  $$
  W_{1,0}(R\,;D) =   \sum_{\substack{I = \{i\} \\ \det{R_I} = 1}} (\nu_i - 1)^2
  + \sum_{\substack{I = \{i\}   \\ \det{R_I} = -1}} (\nu_i + 1)^2
  + \sum_{\substack{I = \{i,j\} \\ \det{R_I} = 1}}  \frac{1}{2}(\nu_i - \nu_j)^2
  + \sum_{\substack{I = \{i,j\} \\ \det{R_I} = -1}} \frac{1}{2}(\nu_i + \nu_j)^2\;.
  $$
\end{theo}

\begin{rem}[On non-distinct entries of $D$]
  If we allow
  $$
  \nu_1 \geq \nu_2 \geq \ldots \geq \nu_n > 0
  $$
  for the entries of $D$, then the $D$- and $R$-invariant
  subspaces $V_i$ are not necessarily coordinate subspaces.
  This produces non-isolated critical points but does not
  change the formula for the critical values.
\end{rem}
In order to compute the global minimizers $R \in \SO(n)$ for the
Cosserat shear-stretch energy $W_{1,0}(R\,;D)$, we have to compare
all the critical values which correspond to the different
partitions and choices of the signs of the determinants
in the statement of \theref{theo:critical_values}. We may, however,
assume that $\det{R_I} = +1$ for all subsets $I$,
see~\cite{Borisov:2016:ORP} for further details.

The following lemma shows that blocks of size two are always
favored \emph{whenever they exist}.
\begin{lem}[Comparison lemma]
  \label{lem:comparison}
  If $\nu_i + \nu_j > 2$ then the difference between
  the critical values of $W_{1,0}(R\,;D)$ corresponding to the
  choice of a size two subset $I = \{i,j\}$ as compared
  to the choice of two size one subsets $\{i\}$, $\{j\}$
  is given by
  $$
  -\frac{1}{2}(\nu_i+\nu_j-2)^2.
  $$
\end{lem}

Let us rewrite $W_{1,0}(R\,;D)$ in a slightly different form in order to
distill the contributions of the size two blocks in the partition.

\begin{cor}
  \label{cor:savings}
For the choices of $\det{R_I} = 1$ there holds
$$
W_{1,0}(R\,;D) = \hsnorm{\sym(RD-\id)}^2 = \sum_{i=1}^n(\nu_i - 1)^2
                  -\frac{1}{2}\sum_{I = \{i,j\}} (\nu_i+\nu_j-2)^2.
$$
\end{cor}
To study the global minimizers for the Cosserat shear-stretch energy
in arbitrary dimension $n \geq 4$, we need to investigate the relative
location of the size two subsets of the partition.

\begin{lem}
  \label{lem:overlap}
  Let $R \in \SO(n)$ be a global minimizer for $W_{1,0}(R\,;D)$.
  Then $R$ cannot contain overlapping size two subsets, i.e.,
  $I = \{i_1,i_4\}$, $J = \{i_2,i_3\}$, with $i_1 < i_2 < i_3 < i_4$.
\end{lem}

We are now ready to state the result in the general $n$-dimensional case.
\begin{theo}
  \label{theo:global_min_nd}
  Let $\nu_1>\nu_2>\ldots \nu_n>0$ be the entries of $D$.
  Let us fix the maximum $k$ for which $\nu_{2k-1} + \nu_{2k} > 2$.
  Any global minimizer $R \in \SO(n)$ corresponds to the
  partition of the form
  $$
  \{1,2\} \sqcup \{3,4\} \sqcup \ldots \sqcup \{2k-1, 2k\}
  \sqcup \{2k+1\} \sqcup \ldots \sqcup \{n\}
  $$
  and the global minimum of $W_{1,0}(R\,;D)$ is given by
\begin{align*}
  W_{1,0}^{\rm red}(D) \eqdef& \min_{R \in \SO(n)}{\hsnorm{\sym(RD - \id)}^2}
  = \sum_{i=1}^n(\nu_i-1)^2 - \frac{1}{2}\sum_{i=1}^k (\nu_{2i-1}+\nu_{2i}-2)^2\\
  =& \frac{1}{2}\sum_{i=1}^k (\nu_{2i-1} - \nu_{2i})^2 + \sum_{i=2k+1}^n (\nu_i-1)^2\;.
\end{align*}
\end{theo}

\begin{rem}
The number of global minimizers in the above theorem is
$2^k$, where $k$ is the number of blocks of size two in
the preceding characterization of a global minimizer as
a block diagonal matrix. All global minimizers are block
diagonal, similar to the previously discussed $n = 3$ case.
\end{rem}
\countres

\addcontentsline{toc}{section}{References}
\bibliographystyle{plain}

{\footnotesize
  \setlength{\bibsep}{1pt}
  \bibliography{fine-2016-grioli}
}

\end{document}